\documentclass[12pt, draftclsnofoot, onecolumn]{IEEEtran}
\usepackage{amsmath}
\usepackage{stmaryrd}
\usepackage{amsfonts}
\usepackage{mathrsfs}
\usepackage{amssymb,color,balance,cite}
\usepackage[dvips]{graphicx}
\usepackage{multicol}
\usepackage{mathtools}
\usepackage{caption}
\usepackage[justification=centering]{caption}
\usepackage{subfigure}
\usepackage{bbm}
\usepackage{multirow}
\usepackage[table]{xcolor}

\newtheorem{remark}{Remark}
\newtheorem{proposition}{Proposition}

 \allowdisplaybreaks[4]

\begin{document}

\title{\Large Angle-Domain Approach for Parameter Estimation in High-Mobility OFDM with Fully/Partly Calibrated Massive ULA
 }

\author{\IEEEauthorblockN{Yinghao Ge, Weile Zhang, Feifei Gao, and Hlaing Minn}
\thanks{Y. Ge and W. Zhang are with the MOE Key Lab for Intelligent Networks and Network Security, Xi'an Jiaotong University, Xi'an, Shaanxi, 710049, China. (email: ge\_yinghao\_jacques@163.com, wlzhang@mail.xjtu.edu.cn).

F. Gao is with the State Key Laboratory of Intelligent Technology and Systems, Tsinghua National Laboratory for Information Science and Technology, Department of Automation, Tsinghua University, Beijing, 100084, China (e-mail: feifeigao@ieee.org).

H. Minn is with the Department of Electrical Engineering, University of Texas at Dallas (e-mail: Hlaing.Minn@utdallas.edu).}
}

 \maketitle

\vspace{-2em}

\begin{abstract}
In this paper, we consider a downlink orthogonal frequency division multiplexing (OFDM) system from a base station to a high-speed train (HST) equipped with fully/partly calibrated massive uniform linear antenna-array (ULA) in wireless environments with abundant scatterers. Multiple Doppler frequency offsets (DFOs) stemming from intensive propagation paths together with transceiver oscillator frequency offset (OFO) result in a fast time-varying frequency-selective channel. We develop an angle domain carrier frequency offset (CFO, general designation for DFO and OFO) estimation approach. A high-resolution beamforming network is designed to separate different DFOs into a set of parallel branches in angle domain such that each branch is mainly affected by a single dominant DFO. Then, a joint estimation algorithm for both maximum DFO and OFO is developed for fully calibrated ULA. Next, its estimation mean square error (MSE) performance is analyzed under inter-subarray mismatches. To  mitigate the detrimental effects of inter-subarray mismatches, we introduce a calibration-oriented beamforming parameter (COBP) and develop the corresponding modified joint estimation algorithm for partly calibrated ULA. Moreover, the Cram\'{e}r-Rao lower bound of CFO estimation is derived. Both theoretical and numerical results are provided to corroborate the proposed method.
\end{abstract}


\section{Introduction}
In a richly scattered high-speed wireless environment, the transmitted signal arrives at the destination after propagating through a number of independent subpaths, with different delays and different angle of arrival (AoA) related Doppler frequency offsets (DFOs). Superposition of these time and frequency shifted versions of transmitted signal at the receiver not only results in the inter-symbol interference (frequency-selective fading), but also leads to a fast time-varying channel, namely time-selective fading. This doubly selective channel fading imposes a significant challenge for high-speed wireless communication~\cite{J_Wu2016Access, F_Hasegawa2017CSCN}. Besides, the transceiver oscillator frequency offset (OFO) naturally exists due to the mismatch of local oscillators. Though orthogonal frequency division multiplexing (OFDM)~\cite{T_Huang2009TVT} is immune to frequency-selective fading, its performance relies heavily on perfect orthogonality among subcarriers, which is quite vulnerable to the carrier frequency offset (CFO, general designation of DFO and OFO)~\cite{S_Ahmed2005TC}. The existence of multiple CFOs will destroy subcarrier orthogonality and cause inter-carrier interference (ICI).
To prevent OFDM from experiencing significant performance degradation, it is crucial to address these multiple CFOs, or the resulting fast time-varying channel.

The most commonly used approach in the current literature to characterize the time variations of channel is basis expansion model (BEM)~\cite{MF_Rabbi2010IET_C, F_Qu2010TWC, H_Hijazi2009TVT, QT_Zhang2006ICC}. The time-varying channel is approximated by the combination of a few basis functions, which greatly reduces the parameters to be estimated. Various candidates of basis functions have been developed, such as complex exponential BEM (CE-BEM)~\cite{F_Qu2010TWC}, polynomial BEM (P-BEM)~\cite{H_Hijazi2009TVT} and Karhunen-Loeve BEM (KL-BEM)~\cite{QT_Zhang2006ICC}. However, the computational burden of BEM methods is still very heavy. Moreover, accurate maximum DFO is necessary so as to determine the minimum order of basis functions, not to mention that KL-BEM further requires accurate channel statistics to compute the basis functions.

Another frequently adopted approach is based on the autocorrelation of the time-varying channel, which could be approximated as the weighted summation of two monochromatic plane waves~\cite{M_Souden2009TSP, F_Bellili2013GLOBECOM, YR_Tsai2009SPL}. In~\cite{M_Souden2009TSP}, channel covariances at different time lags are expressed as a function of Doppler spread factor and OFO, from which a closed-form estimator of Doppler spread factor and OFO is derived. In~\cite{F_Bellili2013GLOBECOM}, the maximum likelihood (ML) estimator is derived and the Doppler spread factor could be estimated via one-dimensional low-cost search. 
However, the studies in~\cite{M_Souden2009TSP, F_Bellili2013GLOBECOM, YR_Tsai2009SPL} are restricted to flat fading channels and circumvent the issues of channel equalization and subsequent data detection.

The doubly selective fading feature of channel is difficult to deal with simply from time or frequency domain. Since this feature originates from multipath propagation associated to different AoAs, it should be more reasonable to resort to angle domain. Some earlier works could be found in~\cite{Y_Zhang2011ICST} and~\cite{W_Guo2013ICSPCC}, where the small-scale uniform circular antenna-array (UCA)~\cite{Y_Zhang2011ICST} and uniform linear antenna-array (ULA)~\cite{W_Guo2013ICSPCC} are adopted  to separate multiple DFOs and eliminate ICI via array beamforming, respectively. However, their work only applies to scenarios with very sparse channels.
Recently, large-scale antenna array has gained growing interest from researchers~\cite{EG_Larsson2014CM, W_Zhang2018TWC, H_Xie2016JSAC}. Profiting from its high spatial resolution, the authors in~\cite{W_Guo2017TVT} made the first attempt to address in angle domain the problem of multiple CFOs under richly scattered high-mobility scenarios. However, the approach in~\cite{W_Guo2017TVT} must exploit a pilot sequence composed of two identical halves in the time domain, which limits its applicability to systems with other pilot sequences, and may not be optimal for channel estimation and subsequent data detection due to sparsity in the frequency domain~\cite{H_Minn2006TC}.

Neither have the authors of~\cite{W_Guo2017TVT} taken into account the fact that it is quite challenging and may not be possible in practice to establish a fully and entirely calibrated large-scale antenna array. Due to various uncontrollable factors such as imperfect time synchronization or communication devices aging, etc., gain and phase mismatches inevitably appear among multiple receive antennas. Thus, for many circumstances involving array signal processing, array calibration is unavoidable prior to exploiting the probable benefits of large-scale antenna array.
Meanwhile, each subarray in a large subarray-based system can be well calibrated, though the calibration of the whole array is quite difficult. A class of partly calibrated subarray-based antenna array has received considerable attention especially in the traditional array signal processing domain~\cite{F_Gao2005SPL, M_Pesavento2002TSP, CMS_See2004TSP}, e.g., rank reduction estimator (RARE) for direction of arrival (DoA) estimation in~\cite{M_Pesavento2002TSP, CMS_See2004TSP}. One of the benefits of dividing the whole uncalibrated antenna array into perfectly calibrated subarrays is that in the case of one or several subarrays being damaged, it is possible to remove or replace the damaged subarrays without affecting the whole antenna array.

Motivated by the above discussions, this paper targets at combating the doubly selective fading channel, by exploiting the high spatial resolution provided by fully/partly calibrated large-scale ULA. First, we design a high-resolution beamforming network to separate the received signal with multiple CFOs into a set of parallel signal branches with single CFO each and develop an estimation algorithm in the case of fully calibrated ULA to jointly acquire maximum DFO, OFO and channel. After that the conventional CFO compensation and maximum-ratio-combining (MRC)-based data detection are performed. Next, the CFO estimation mean square error (MSE) performance analysis unveils its incapability in dealing with inter-subarray mismatches. In view of this, the above algorithm is further modified by introducing a calibration-oriented beamforming parameter (COBP), making it applicable to partly calibrated ULA.
In summary, the main contribution of this paper can be described as follows:
\begin{itemize}
  \item The frequency synchronization and channel estimation problem in high-mobility scenarios with both DFO and OFO is addressed, whether the ULA at the receiver is fully or partly calibrated. By taking into account the inter-subarray gain and phase mismatches, our system model represents a more realistic scenario and the introduction of COBP effectively remedies the detrimental effects of those array mismatches.
  \item By eliminating the necessity of exploiting the two-halves pilot as in~\cite{W_Guo2017TVT}, the proposed joint estimation algorithms can be implemented into many practical communication systems without incurring incompatibility problem.
      Moreover, as the proposed algorithms can be applied to systems with optimized pilot sequences, its performance can be superior to~\cite{W_Guo2017TVT}.
  \item An alternative solution based on Taylor series expansion is developed to address the extremely time-consuming two-dimensional grid-search, thereby prominently reducing the computational complexity of the proposed joint estimation algorithms.
  \item The MSE performance analysis justifies the necessity of introducing COBP in the presence of array mismatches, and reveals the twofold relationship between the estimation MSE of DFO and that of OFO in the case of fully calibrated ULA. In addition, the Cram\'{e}r-Rao Bound (CRB) for joint DFO and OFO estimation is derived in order to theoretically assess the performance of the proposed algorithms.
\end{itemize}

The remainder of this paper is organized as follows. The system model is described in Section II. Section III presents the joint estimation algorithm designed for the fully calibrated ULA. Section IV first applies the joint estimation algorithm in Section III to the partly calibrated case and analyzes the MSE performance, and then extends the joint estimation algorithm to the partly calibrated ULA. The CRB is derived in Section V. Simulation results are provided in Section VI. Section VII draws the conclusion of this paper. Part of this paper has appeared in a conference paper~\cite{Y_GE2017VTCSpring}.

\textit{Notations:} Superscripts $(\cdot)^*$, $(\cdot)^T$, $(\cdot)^H$, $(\cdot)^\dag$ and $E{\{\cdot\}}$
represent conjugate, transpose, conjugated transpose, pseudo-inverse and expectation, respectively; ${\rm j}=\sqrt{-1}$ is the imaginary unit; $\mathbf{\Re} \{\cdot\}$ and $\mathbf{\Im} \{\cdot\}$ denote the real and imaginary part, and $\|\cdot\|_{2}$ denotes the Euclidean norm of a vector or Frobenius norm of a matrix. $\operatorname{diag}(\bf x)$ is a diagonal matrix with the vector $\mathbf{x}$ as its diagonal elements, and $\operatorname{blkdiag}(\cdot)$ represents a block diagonal matrix; $\operatorname{tr}(\cdot)$ denotes the trace operator, $\lambda_{\textrm{min}}(\cdot)$ and ${\mathbf{v}}_{\min }(\cdot)$ return respectively the minimum eigenvalue and corresponding eigenvector of a positive semi-definite matrix.
$\otimes$ and $\odot$ stand for the Kronecker product and Schur-Hadamard product (element-wise product), respectively. ${\mathbf{I}_N}$ is the $N\times N$ identity matrix, ${\bf 1}$ and ${\bf 0}$ represent respectively an all-one and all-zero matrix with appropriate dimension.

\vspace{-0.6em}
\section{System Model}
Consider the OFDM downlink transmission in a high-mobility scenario where the signal transmitted from base station (BS) arrives at the high-speed train (HST) along a number of independent subpaths. The HST is equipped with a fully or partly calibrated massive ULA for decoding data from BS and then delivering to target users. We will describe in this section the complete system model with the partly calibrated ULA, while the fully calibrated ULA can be regarded as a special case.

\vspace{-0.8em}
\subsection{Actual Steering Vector for the Partly Calibrated Antenna Array}
Consider that the terminal HST is equipped with a massive ULA deployed along the direction of HST motion. The ULA consists of $M$ antennas which can be evenly decomposed into $K$ subarrays, each with $J = M/K$ elements. We suppose that each subarray is well calibrated while the calibration of the whole array is imperfect due to AoA-independent inter-subarray gain and phase mismatches~\cite{CMS_See2004TSP}.
Denote $\boldsymbol{\varepsilon}$ as the calibration error parameter. Then, the actual steering vector towards direction $\theta$ could be written as~\cite{CMS_See2004TSP}
\begin{align}
\mathbf{a}(\theta, \boldsymbol \varepsilon)=\mathbf{V}(\theta)\boldsymbol \alpha(\boldsymbol \varepsilon),
\end{align}
where $\mathbf{V}(\theta) \!=\! \operatorname{blkdiag}\left({{\mathbf{v}}_{1}}(\theta), \!\ {{\mathbf{v}}_{2}}(\theta), \!\ \cdots\!, \!\ {{\mathbf{v}}_{K}}(\theta)\right) $ and $\boldsymbol{\alpha}(\boldsymbol{\varepsilon}) \!=\! {{\left[ \begin{matrix}
   {{\alpha}_{1}},\!\! & {{\alpha }_{2}},\!\! & \cdots\!,\!\!  & {{\alpha }_{K}}  \\
\end{matrix}\right]}^{T}}$. Here, $\mathbf{V}(\theta)$ \\ is an $M\times K$ block-diagonal matrix whose non-zero elements ${{\left[ \begin{matrix}
   \mathbf{v}_{1}^{T}(\theta), & \mathbf{v}_{2}^{T}(\theta), & \cdots\!,  & \mathbf{v}_{K}^{T}(\theta)  \\
\end{matrix}\right]}^{T}}$ correspond to the array response vector of an $M$-elements fully calibrated ULA $\mathbf{a}(\theta)$, and $\boldsymbol{\alpha}(\boldsymbol{\varepsilon})$ is the $K\times 1$ complex vector characterizing inter-subarray gain and phase mismatches, with $\alpha_k$ being the gain and phase mismatch of the $k\rm{th}$ subarray.

Further denote $d$ as the antenna spacing and $\lambda$ as the carrier wavelength. Then, the $r$th element of $\mathbf{a}(\theta, \boldsymbol \varepsilon)$ is given by ${{a}_{r}}\left( {{\theta }}, \boldsymbol{\varepsilon } \right) = {{\alpha }_{k}}{{e}^{\text{j}2\chi\left( r-1 \right)\cos {{\theta }}}}$, where $\chi = \pi \tilde{d} $, $\tilde{d} = \frac{d}{\lambda}$ and $k$ denotes the subarray index to which the $r$th antenna belongs.

\vspace{-0.9em}
\subsection{Fast Time-varying Channel Model and Received Signal at Partly Calibrated Antenna Array}
\begin{figure}[t]
\setlength{\abovecaptionskip}{-0.5cm}
\setlength{\belowcaptionskip}{-1.0cm}
\begin{center}
\includegraphics[width=100mm]{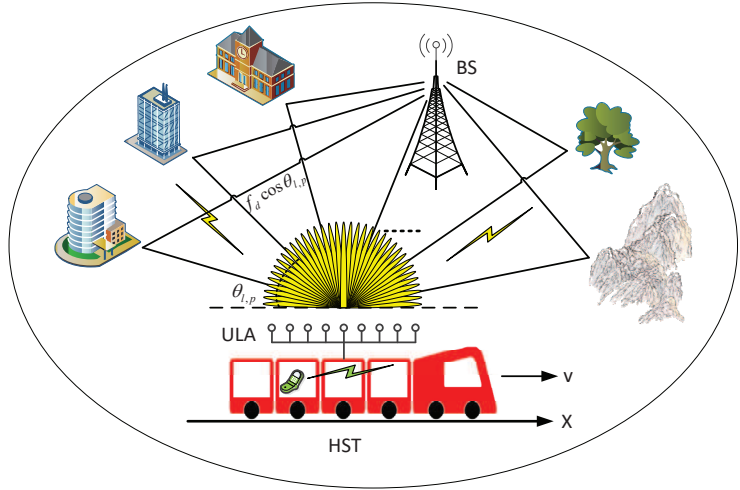}
\end{center}
\caption{ Illustration of the richly scattered HST scenario with multi-branch beamforming towards pre-fixed directions \\ (only a few subpaths are shown as example). }
\end{figure}

The considered scenario where the fast-moving terminal is surrounded by plentiful scatterers, as illustrated in Fig. 1, could be fairly characterized by an established Jakes' channel model~\cite{bWC_Jakes1994, YR_Zheng2003TC}. The channel between the BS and the $r$th antenna is modeled as $L$ taps $l\!=\!1,2,\cdots\!,L $, each tap composed of $P \!\gg\! 1$ separable subpaths with index $p\!=\!1,2,\cdots\!,P $, which could be identified by its unique AoA $\theta_{l,p} \sim U\left(0,2\pi \right)$ and associated complex gain $g_{l,p} \sim \mathcal{CN}\left( 0, {\sigma_{l}^{2}}/P \right)$. Here, $\{\sigma_{l}^{2}, l\!=\!1,2,\cdots\!, L\}$ models the channel power delay profile (PDP). We assume $\sum\nolimits_{l=1}^{L}{\sigma_{l}^{2}} = 1$ such that the total average channel gain per receive antenna is normalized. The channel impulse response of the $p$th subpath at the $l$th tap can be written as $\mathbf{h}\left( {{\theta }_{l,p}} \right)\in {{\mathbb{C}}^{L\times 1}}$, whose $l'$th element is given by ${{g}_{l, p}}\delta \left( l'-{l} \right)$.

Denote $\xi$ as the normalized OFO (nOFO) relative to the subcarrier spacing ${{f}_{s}}$. Denote $f_d$ as the normalized maximum DFO (nDFOmax) defined as the ratio of maximum Doppler shift $\frac{\upsilon }{\lambda }$ and ${{f}_{s}}$, where $\upsilon$ refers to HST velocity. Thus, the normalized DFO (nDFO) and the effective superimposed CFO for the $p$th subpath at the $l$th tap are determined by ${{f}_{l, p}}={{f}_{d}}\cos {{\theta }_{l, p}}$ and ${{\varphi }_{l,p}}={{f}_{l,p}}+\xi $, respectively.

Consider one OFDM frame consisting of $N_b$ OFDM blocks, where the first block serves as pilot and the rest is reserved for data transmission. Note that for Jakes' channel model, each subpath has independent attenuation, phase and AoA (thus different DFO) and we assume that $g_{l,p}$, $\theta_{l,p}$ and $f_{l,p}$ may differ among different frames but remain constant over one OFDM frame.

Denote ${{\mathbf{x}}_{m}} \!=\! {{\left[ \begin{matrix}
   {{x}_{m, 0}}, & {{x}_{m, 1}}, & \cdots\!,  & {{x}_{m, N-1}}  \\
\end{matrix} \right]}^{T}}$ as the frequency domain pilot ($m\!=\!1$) or data symbols ($m\!>\!1$) in the $m$th block, with $N$ being the number of subcarriers. Define $\mathbf{F}$ as the $N\!\times\! N$ unitary Discrete Fourier Transform (DFT) matrix, with $\frac{1}{\sqrt{N}}{{e}^{-\text{j}2{\rm\pi} \frac{\left( k-1 \right)\left( n-1 \right)}{N}}}$ as its $(k,n)$-th entry and ${{\mathbf{F}}_{L}}$ the submatrix consisting of its first $L$ columns. Then, the time domain samples can be obtained by applying an $N$-point inverse DFT on ${\mathbf{x}}_{m}$, i.e., ${{\mathbf{s}}_{m}} \!=\! {\mathbf{F}}^H {\mathbf{x}}_{m}$. Appending the cyclic prefix (CP) of length $N_{\mathrm{cp}}$ to the time domain samples ${{\mathbf{s}}_{m}}$ yields ${{\mathbf{s}}_{m, \mathrm{cp}}}$. The existence of CP turns the linear convolution between ${{\mathbf{s}}_{m, \mathrm{cp}}}$ and $\mathbf{h}\left( {{\theta }_{l,p}} \right)$ into circular convolution between ${{\mathbf{s}}_{m}}$ and $\mathbf{h}\left( {{\theta }_{l,p}} \right)$, i.e., $\mathbf{s}_m \circledast \mathbf{h}(\theta_{l,p})$.
Therefore, the signal in the $m$th block (after CP removal) received from the $p$th subpath at the $l$th tap can be expressed as the following $N\times M$ matrix
\begin{align}
{{\mathbf{Y}}_{m}(\theta_{l,p})} = {{{\eta }_{m}}\left( {{\varphi }_{l,p}} \right) \mathbf{E}\left( {{\varphi }_{l,p}} \right) \left(\mathbf{s}_m \circledast \mathbf{h}(\theta_{l,p})\right) {{\mathbf{a}}^{T}}\left( {{\theta }_{l, p}}, \boldsymbol{\varepsilon } \right)},
\end{align}
where ${{\eta }_{m}}\left( {{\varphi }_{l,p}} \right)={{e}^{\text{j}2{\rm\pi} {{\varphi }_{l,p}}\frac{\left( m-1 \right)\left( N+{{N}_{\mathrm{cp}}} \right)}{N}}}$ is the accumulative phase shift of the $m$th block induced by ${\varphi }_{l,p}$, and $\mathbf{E}\left( {{\varphi }_{l,p}} \right)=\operatorname{diag}\left( 1,\ {{e}^{\text{j}2{\rm\pi} {{\varphi }_{l,p}}\frac{1}{N}}},\ \cdots\!, \ {{e}^{\text{j}2{\rm\pi} {{\varphi }_{l,p}}\frac{N-1}{N}}} \right)$ represents the phase rotation inside one OFDM block. Note that we assume perfect time synchronization between the transceivers.

Considering that the circular convolution in the time domain corresponds to the pointwise multiplication in the frequency domain, we have
\begin{align}
\mathbf{s}_m \circledast \mathbf{h}(\theta_{l,p}) &= \mathbf{F}^H \mathbf{F} \left(\mathbf{s}_m \circledast \mathbf{h}(\theta_{l,p})\right) = \mathbf{F}^H  \operatorname{diag}(\mathbf{F} \mathbf{s}_m) \sqrt{N} \mathbf{F}_L \mathbf{h}(\theta_{l,p}) = {{\mathbf{B}}_{m}}\mathbf{h}\left( {{\theta }_{l,p}} \right),
\end{align}
where ${{\mathbf{B}}_{m}} \!=\! \sqrt{N}{{\mathbf{F}}^{H}} \operatorname{diag} \left( {{\mathbf{x}}_{m}} \right) {{\mathbf{F}}_{L}}$.
As a result, the total received signal in the $m$th block can be finally expressed as
\begin{align} \label{ReceivedSignal}
{{\mathbf{Y}}_{m}} = \sum\limits_{l=1}^{L}{\sum\limits_{p=1}^{P}{{{\eta }_{m}}\left( {{\varphi }_{l,p}} \right)\mathbf{E}\left( {{\varphi }_{l,p}} \right){{\mathbf{B}}_{m}}\mathbf{h}\left( {{\theta }_{l,p}} \right){{\mathbf{a}}^{T}}\left( {{\theta }_{l, p}}, \boldsymbol{\varepsilon } \right)}} + {{\mathbf{W}}_{m}},
\end{align}
where ${{\mathbf{W}}_{m}}$ is the zero-mean complex additive white Gaussian noise (AWGN) in the $m$th block at the receive antenna array with $E\{{{\mathbf{W}}_{m}}{{\mathbf{W}}_{m}^H}\} \!=\! M{\sigma_{\mathrm{n}}^{2}}{\mathbf{I}_L}$. Here, $\sigma_{\mathrm{n}}^{2}$ denotes the noise power.

\vspace{-0.6em}
\section{Proposed joint estimation algorithm for fully calibrated ULA}
The integrated receiving procedure with fully calibrated massive ULA will be elaborated in this section, and the diagram of this procedure is illustrated in Fig. 2. First, a high-resolution beamforming network is designed to separate the received signal into $Q$ parallel branches, each of which is mainly affected by a single CFO. Then, the CFO and channel are jointly estimated, using all the $Q$ beamforming branches. Next, conventional CFO compensation techniques could be performed for each branch. Finally, MRC is utilized to recover the transmitted data symbols.

\begin{figure}[t]
\setlength{\abovecaptionskip}{-0.5cm}
\setlength{\belowcaptionskip}{-1cm}
\begin{center}
\includegraphics[width=100mm]{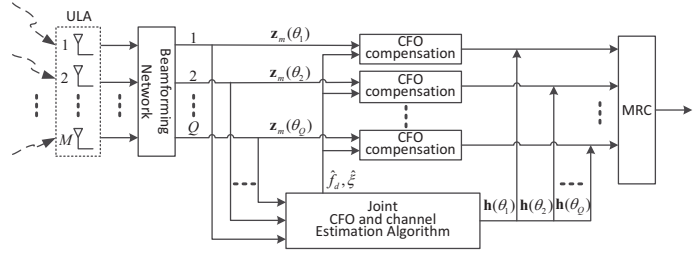}
\end{center}
\caption{ Diagram of the receiver design in the case of fully calibrated ULA.}
\end{figure}

\vspace{-0.6em}
\subsection{Beamforming Network}
From (\ref{ReceivedSignal}), the received pilot signal without considering inter-subarray mismatches is given by (the subscript $m=1$ denoting the pilot block is omitted for brevity hereafter)
\begin{align}
\mathbf{Y}=\sum\limits_{l=1}^{L}{\sum\limits_{p=1}^{P}{\mathbf{E}\left( {{\varphi }_{l,p}} \right)\mathbf{Bh}\left( {{\theta }_{l,p}} \right){{\mathbf{a}}^{T}}\left( {{\theta }_{l, p}} \right)}}+\mathbf{W}.
\end{align}

Since the multiple CFOs are related to different AoAs, the difficulty could be alleviated if we can separate signals of different AoAs through a high-resolution beamforming network. Owing to sufficient number of antennas, the steering vectors of a fully calibrated massive ULA pointing to any two different directions are nearly orthogonal, i.e., ${{\mathbf{a}}^{H}}\left( {{\theta }_{1}} \right)\mathbf{a}\left( {{\theta }_{2}} \right) \approx 0, {{\theta }_{1}}\ne {{\theta }_{2}}$. Such orthogonality helps eliminate the inter-direction interference, thereby enabling the steering vector a very simple but effective candidate beamformer.
Since signals may come from any direction in the space due to rich scatterers around the moving HST, it is more reasonable to directly perform beamforming for a range of pre-fixed $\theta$ without estimating the AoAs previously. How to determine the range of $\theta$ will be discussed later.

Define the beamformer $\mathbf{b}\left( \theta  \right) = \frac{1}{M} \mathbf{a}\left( \theta  \right) $. The received signal $\mathbf{z}\left( \theta  \right) = \mathbf{Y}{{\mathbf{b}}^{\text{*}}}\left( \theta  \right)$ is expressed as
\begin{align} \label{SIN}
\mathbf{z}\left( \theta  \right) = \underbrace{\sum\limits_{l,p, {{\theta }_{l,p}}
 =\theta }{\mathbf{E}\left( {{\varphi }_{l,p}} \right)\mathbf{Bh}\left( {{\theta }_{l,p}} \right)}}_{\mathrm{desired \ signal}} +\underbrace{\sum\limits_{l', p', {{\theta }_{l',p'}}\ne \theta }{\mathbf{E}\left( {{\varphi }_{l',p'}} \right)\mathbf{Bh}\left( {{\theta }_{l',p'}} \right){{\mathbf{a}}^{T}}\left( {{\theta }_{l', p'}} \right){{\mathbf{b}}^{\text{*}}}\left( \theta  \right)}}_{\mathrm{interference}} +\underbrace{\mathbf{W}{{\mathbf{b}}^{\text{*}}}\left( \theta  \right)}_{\mathrm{noise}},
\end{align}
where the first term is the desired signal relative to the interested direction $\theta$, while the second and third terms represent the interference from other directions and the noise after beamforming, respectively. The sufficient spatial dimension provided by massive ULA creates a high-resolution beamformer, which only allows signals arriving from the interested direction $\theta$ to pass through and mitigates prominently the interference from other directions. This makes the second term negligible. By ignoring the interference, we arrive at
\begin{align}
\mathbf{z}\left( \theta  \right) = \mathbf{E}\left( {{f}_{d}}\cos \theta +\xi  \right)\mathbf{B}\mathbbm{h} \left( \theta  \right)+\mathbf{\tilde{w}}\left( \theta \right),
\end{align}
where $\mathbbm{h}\left( \theta  \right)=\sum\nolimits_{{{\theta }_{l,p}}=\theta }{\mathbf{h}\left( {{\theta }_{l,p}} \right)}$
and $\mathbf{\tilde{w}}\left( \theta \right) =\mathbf{W}{{\mathbf{b}}^{\text{*}}}\left( \theta  \right)$.
Note that in the case of no signals arriving from direction $\theta$, $\mathbbm{h}\left( \theta  \right)$ equals $\mathbf{0}$ and thus $\mathbf{z}\left( \theta  \right)$ only comprises noise and weak interference.

Now, we will discuss how to determine some critical parameters for the beamforming network, such as the range of direction $\theta$ and antenna spacing $d$.
First, in the considered richly scattered scenario, the signals may come from all directions in the entire space. The cone-shaped beam pattern of ULA (as shown in Fig. 3 of~\cite{W_Guo2017TVT}) can guarantee that all the signals incorporated by the beam towards $\theta$ are mainly affected by the same CFO ${{f}_{d}}\cos \theta +\xi $. Besides, the adoption of ULA only requires to design the beamforming network along the dimension of AoA for receiving all signals dispersed in the whole space. Therefore, ULA is preferable to deal with the considered scenario.
Moreover, the ULA cannot differentiate two symmetric AoAs $\theta$ and $360^\circ\!-\!\theta$, making it sufficient to perform beamforming within the range of $(0^\circ,180^\circ)$.

Second, the antenna spacing $d$ optimizing beamforming resolution without incurring aliasing is $d=\frac{\lambda}{2}$. However, this cannot avoid the aliasing between $0^\circ$ (corresponding nDFO $f_d$) and $180^\circ $ (corresponding nDFO $-f_d$), which will bring inconvenience for CFO compensation. Hence, a tradeoff between beamforming resolution and aliasing avoidance needs to be taken.

Third, since an $M$-elements ULA provides at most $M$ degrees of freedom (DoF), it is sufficient to perform beamforming towards $M$ distinct directions, which could either be evenly selected between $(0^\circ,180^\circ)$, or drawn from ${\mathcal{D}}_{{\rm{IFFT}}} \!=\! \left\{ {{\vartheta}_{1}}, {{\vartheta}_{2}}, \cdots\!, {{\vartheta}_{M}} \right\}$ with ${{\vartheta}_{r}} \!=\! {\rm{arccos}}\left( \frac{\rm{\pi}}{\chi} \left(\frac{r-1}{M} \!-\! \frac{1}{2} \right)\right)$~\cite{L_You2015TWC}. Here, $\frac{1}{\sqrt{M}} {{\left[ \begin{matrix}
   \mathbf{a}(\vartheta_1), & \mathbf{a}(\vartheta_2), & \cdots \!, & \mathbf{a}(\vartheta_M) \\
\end{matrix} \right]}}$ in fact constitutes the column-permuted normalized inverse DFT matrix and thus the beamforming can be efficiently achieved via FFT operation. However, as $\tilde{d}<0.5$, we have $\left| \frac{\rm{\pi}}{\chi} \left(\frac{r-1}{M} \!-\! \frac{1}{2} \right) \right| > 1$ for $r \!<\! \left(\frac{1}{2} \!-\! \tilde{d} \right)M\!+\!1$ or $r \!>\! \left(\frac{1}{2} \!+\! \tilde{d} \right)M\!+\!1$ and thereby no corresponding real physical angles ${{\vartheta}_{r}}$ can be found. That is to say, the beamforming can only be performed towards $Q\!<\!M$ directions at $\tilde{d}\!<\!0.5$. As a result, by exploiting FFT operation, we benefit from computational efficiency and perfect orthogonality among different beamformers at the cost of slightly sacrificing some DoF.

\vspace{-0.6em}
\subsection{Joint Estimation Algorithm with Fully Calibrated ULA}
The beamforming decomposes the received signal into $Q$ parallel branches, each of which is affected by a single CFO. Assuming perfect interference elimination, we can find the estimates of nDFOmax ${\hat{f}}_{d}$, nOFO $\hat{\xi }$ and channel $\mathbbm{\hat{h}}\left( \theta_{q}  \right)$ for the $q$th branch by solving the minimization problem below
\begin{align} \label{OptimPb0}
\left\{{\hat{f}}_{d}, \hat{\xi}, \mathbbm{\hat{h}}\left( \theta_{q}  \right)\right\} = \arg \underset{\left\{{\tilde{f}}_{d}, \tilde{\xi}, \mathbbm{\tilde{h}}\left( \theta_{q}  \right)\right\}}{\mathop{\min }}\,\left\| \mathbf{z}\left( \theta_{q}  \right)-\mathbf{E}\left( \tilde{\varphi}_{q} \right)\mathbf{B}\mathbbm{\tilde{h}} \left( \theta_{q}  \right) \right\|_{2}^{2},
\end{align}
where $\tilde{\varphi}_{q} = {{\tilde{f}}_{d}}\cos \theta_{q} + \tilde{\xi}$. For the given trial value pair ${{\tilde{f}}_{d}}$ and $\tilde{\xi }$, the ML estimator of $\mathbbm{\hat{h}}(\theta_{q})$ minimizing (\ref{OptimPb0}) is given by $\mathbbm{\hat{h}}\left( \theta_{q}  \right)={{\mathbf{B}}^{\dagger }}{{\mathbf{E}}^{H}}\left( \tilde{\varphi}_{q} \right)\mathbf{z}\left( \theta_{q}  \right)$. Let ${{\mathbf{P}}_{\mathbf{B}}}=\mathbf{B}{{\mathbf{B}}^{\dagger }}$ represent the orthogonal projection operator onto the subspace spanned by the columns of $\mathbf{B}$ and $\mathbf{P}_{\mathbf{B}}^{\bot }={{\mathbf{I}}_{N}}-{{\mathbf{P}}_{\mathbf{B}}}$. Then by substituting $\mathbbm{\tilde{h}}\left( \theta_{q}  \right)$ with its ML estimator, (\ref{OptimPb0}) is reduced to
%
\begin{align}\label{OptimPb_SingleBeam}
\left\{{\hat{f}}_{d}, \hat{\xi}\right\} = \arg \underset{\left\{{{\tilde{f}}}_{d}, \tilde{\xi }\right\}}{\mathop{\min }}\,{g\left( {{{\tilde{f}}}_{d}}, \tilde{\xi }, {{\theta }_{q}} \right)},
\end{align}
where $g\left( {{{\tilde{f}}}_{d}}, \tilde{\xi }, {{\theta }_{q}} \right) = \left\| \mathbf{P}_{\mathbf{B}}^{\bot }{{\mathbf{E}}^{H}}\left( \tilde{\varphi}_{q} \right)\mathbf{z}\left( \theta_{q}  \right) \right\|_{2}^{2}$.
It should be pointed out that an estimate of superimposed CFO $\hat{\varphi}_q$ can be acquired by solving (\ref{OptimPb_SingleBeam}). However, there are infinite combinations of $\hat{f}_d$ and $\hat{\xi}$ satisfying $\hat{\varphi}_q \!=\! \hat{f}_d \cos\theta_q \!+\! \hat{\xi}$. In other words, ambiguity exists between DFO and OFO estimation if only one beamforming branch is used. Since nDFOmax and nOFO are the same for all branches~\cite{C_Tepedelenlioglu2001TVT}, we can employ simultaneously all the beamforming branches to eliminate such estimation ambiguity. Namely, nDFOmax and nOFO can be jointly estimated from
\begin{align}\label{OptimPb1}
\left\{{{\hat{f}}}_{d}, \hat{\xi }\right\}=\arg \underset{\left\{{{\tilde{f}}}_{d}, \tilde{\xi}\right\}}{\mathop{\min }}\, {g\left( {{{\tilde{f}}}_{d}}, \tilde{\xi } \right)},
\end{align}
where $g\left( {{{\tilde{f}}}_{d}}, \tilde{\xi } \right) = \sum\nolimits_{q=1}^{Q}{g\left( {{{\tilde{f}}}_{d}}, \tilde{\xi }, {{\theta }_{q}} \right)}$.
The equivalent channel of the $q$th beamforming branch is given by $\mathbbm{\hat{h}}\left( {{\theta }_{q}} \right)={{\mathbf{B}}^{\dagger }}{{\mathbf{E}}^{H}}\left( {{{\hat{f}}}_{d}}\cos {{\theta }_{q}}+\hat{\xi } \right)\mathbf{z}\left( {{\theta }_{q}} \right)$.

Directly solving (\ref{OptimPb1}) necessitates the exhaustive two-dimensional grid-searching. Instead, we solve (\ref{OptimPb1}) efficiently and iteratively via Newton's method~\cite{S_Boyd2004}.
Let $\hat{f}_{d}^{\left( i-1 \right)}$ and $\hat{\xi}^{\left( i-1 \right)}$ represent the estimation of ${f}_{d}$ and that of $\xi$ in the $\left( i-1 \right)$th iteration, respectively. Moreover, define $\hat{\varphi}_{q}^{\left( i-1 \right)} \!=\! \hat{f}_{d}^{\left( i-1 \right)}\cos{{\theta }_{q}} + \hat{\xi}^{\left( i-1 \right)}$, $\tilde{\varphi}_{q}^{\left( i \right)} \!=\! \tilde{f}_{d}^{\left( i \right)}\cos{{\theta }_{q}} + \tilde{\xi}^{\left( i \right)}$ and $\Delta {\tilde{\varphi}^{\left( i\right)}_{q}} \!=\! \Delta {\tilde{f}^{\left( i\right)}_{d}}\cos{{\theta }_{q}} + \Delta \tilde{\xi}^{\left( i\right)}$, where $\Delta {\tilde{f}_{d}^{\left( i \right)}} \!=\! {\tilde{f}^{\left( i\right)}_{d}} \\ \!-\! \hat{f}_{d}^{\left( i-1 \right)}$ and $\Delta \tilde{\xi}^{\left( i \right)} \!=\! \tilde{\xi}^{\left( i\right)} \!-\! \hat{\xi}^{\left( i-1 \right)}$ denote the trial value pair of residual nDFOmax and residual nOFO in the $i$th iteration, respectively.

Define $\mathbf{\hat{z}}^{\left( i \right)}\left( {{\theta }_{q}} \right)={{\mathbf{E}}^{H}}\left( \hat{\varphi}_{q}^{\left( i-1 \right)} \right)\mathbf{z}\left( {{\theta }_{q}} \right)$ and $\mathbf{D}={\mathrm{j}}\frac{2{\rm \pi}}{N} \operatorname{diag}\left( 0, 1, \cdots\!, N-1 \right)$. Then, with Taylor series expansion $\mathbf{E}\left( {{\tilde{\varphi}}}^{\left( i\right)}_{q} \right) \approx \mathbf{E}\left( \hat{\varphi}_{q}^{\left( i-1 \right)} \right) \left( {{\mathbf{I}}_{N}}+ \Delta {{\tilde{\varphi}}}^{\left( i\right)}_{q} \mathbf{D}+\frac{1}{2}{{\left( \Delta {{\tilde{\varphi}}}^{\left( i\right)}_{q} \right)}^{2}}{{\mathbf{D}}^{2}} \right)$,
$g\left( {{{\tilde{f}}}_{d}^{(i)}}, {\tilde{\xi}}^{(i)}, {{\theta }_{q}} \right)= \left\| \mathbf{P}_{\mathbf{B}}^{\bot } {\mathbf{E}}^{H}\left( {{\tilde{\varphi}}}^{\left( i\right)}_{q} \right) \mathbf{z}\left( {{\theta }_{q}} \right) \right\|_{2}^{2}$ could be approximated as
\begin{align}
g\left( {{{\tilde{f}}}_{d}^{(i)}}, {\tilde{\xi}}^{(i)}, {{\theta }_{q}} \right)
  \approx {{T}_{0}^{\left( i \right)}}\left( {{\theta }_{q}} \right)+ \Delta \tilde{\varphi}_{q}^{\left( i \right)} {{T}_{1}^{\left( i \right)}}\left( {{\theta }_{q}} \right) +{{\left( \Delta \tilde{\varphi}_{q}^{\left( i \right)} \right)}^{2}}{{T}_{2}^{\left( i \right)}}\left( {{\theta }_{q}} \right),
\end{align}
where
\begin{align*}
{{T}_{0}^{\left( i \right)}}\left( {{\theta }_{q}} \right)&={{{\mathbf{\hat{z}}}}^{{\left( i \right)}H}}\left( {{\theta }_{q}} \right)\mathbf{P}_{\mathbf{B}}^{\bot }\mathbf{\hat{z}}^{\left( i \right)}\left( {{\theta }_{q}} \right), \
{{T}_{1}^{\left( i \right)}}\left( {{\theta }_{q}} \right) = 2\Re \left\{ {{{\mathbf{\hat{z}}}}^{{\left( i \right)}H}}\left( {{\theta }_{q}} \right)\mathbf{DP}_{\mathbf{B}}^{\bot }\mathbf{\hat{z}}^{\left( i \right)}\left( {{\theta }_{q}} \right) \right\}, \nonumber \\
{{T}_{2}^{\left( i \right)}}\left( {{\theta }_{q}} \right) &= \Re \left\{ {{{\mathbf{\hat{z}}}}^{{\left( i \right)}H}}\left( {{\theta }_{q}} \right) {{\mathbf{D}}^{2}}\mathbf{P}_{\mathbf{B}}^{\bot } \mathbf{\hat{z}}^{\left( i \right)}\left( {{\theta }_{q}} \right) \right\} + {{{\mathbf{\hat{z}}}}^{{\left( i \right)}H}}\left( {{\theta }_{q}} \right) \mathbf{DP}_{\mathbf{B}}^{\bot }{{\mathbf{D}}^{H}} \mathbf{\hat{z}}^{\left( i \right)}\left( {{\theta }_{q}} \right).
\end{align*}

Therefore, we obtain
\begin{align} \label{QuadraticFun1}
g\left( {{{\tilde{f}}}_{d}^{(i)}}, {\tilde{\xi}}^{(i)} \right)
 \approx {{t}_{0}^{\left( i \right)}}+{{t}_{11}^{\left( i \right)}} \Delta {{\tilde f}_{d}^{\left( i \right)}} + {{t}_{12}^{\left( i \right)}} \Delta \tilde\xi^{\left( i \right)} + {{t}_{21}^{\left( i \right)}} {{\left( \Delta {{\tilde f}_{d}^{\left( i \right)}} \right)}^{2}} + {{t}_{22}^{\left( i \right)}} \Delta {{\tilde f}_{d}^{\left( i \right)}}\Delta \tilde \xi^{\left( i \right)} + {{t}_{23}^{\left( i \right)}} {{\left( \Delta \tilde \xi^{\left( i \right)}  \right)}^{2}},
\end{align}
where
\begin{align*}
& {{t}_{0}^{\left( i \right)}}=\sum\nolimits_{q=1}^{Q}{{{T}_{0}^{\left( i \right)}}\left( {{\theta }_{q}} \right)},\ \ {{t}_{11}^{\left( i \right)}}=\sum\nolimits_{q=1}^{Q}{\cos{{\theta }_{q}}{{T}_{1}^{\left( i \right)}}\left( {{\theta }_{q}} \right)}, \ \ {{t}_{12}^{\left( i \right)}}=\sum\nolimits_{q=1}^{Q}{{{T}_{1}^{\left( i \right)}}\left( {{\theta }_{q}} \right)\ } \nonumber \\
& {{t}_{21}^{\left( i \right)}}=\sum\nolimits_{q=1}^{Q}{{\cos}^{2}{{\theta }_{q}}{{T}_{2}^{\left( i \right)}}\left( {{\theta }_{q}} \right)}, \ \ {{t}_{22}^{\left( i \right)}}=\sum\nolimits_{q=1}^{Q}{2\cos{{\theta }_{q}}{{T}_{2}^{\left( i \right)}}\left( {{\theta }_{q}} \right)},\ \ {{t}_{23}^{\left( i \right)}}=\sum\nolimits_{q=1}^{Q}{{{T}_{2}^{\left( i \right)}}\left( {{\theta }_{q}} \right)}.
\end{align*}

By setting the first-order gradients of (\ref{QuadraticFun1}) with respect to $\Delta {\tilde{f}_{d}^{\left( i \right)}}$ and $\Delta \tilde{\xi}^{\left( i \right)} $ equal zeros, the optimal residual nDFOmax and that of residual nOFO in the $i$th iteration are given by
\begin{align} \label{CFOSolution}
\left[ \begin{matrix}
   \Delta \hat{f}_{d}^{\left( i \right)}  \\
   \Delta \hat{\xi}^{\left( i \right)}  \\
\end{matrix} \right]=-{{\left[ \begin{matrix}
   2{{t}_{21}^{\left( i \right)}} & {{t}_{22}^{\left( i \right)}}  \\
   {{t}_{22}^{\left( i \right)}} & 2{{t}_{23}^{\left( i \right)}}  \\
\end{matrix} \right]}^{-1}}\left[ \begin{matrix}
   {{t}_{11}^{\left( i \right)}}  \\
   {{t}_{12}^{\left( i \right)}}  \\
\end{matrix} \right].
\end{align}

Thus, the CFO estimation in the $i$th iteration could be accordingly updated as $\hat{f}_{d}^{\left( i \right)} \!=\! \hat{f}_{d}^{\left( i-1 \right)} \!+\! \Delta \hat{f}_{d}^{\left( i \right)}$ and $\hat{\xi}^{\left( i \right)} \!=\! \hat{\xi}^{\left( i-1 \right)} \!+\! \Delta \hat{\xi}^{\left( i \right)}$.
Note that $\hat{f}_{d}^{\left( 0 \right)} \!=\! 0$ and $\hat{\xi}^{\left( 0 \right)} \!=\! 0$ are used for initialization and the convergence to the local optimum of the Newton's method is well proved in~\cite{S_Boyd2004}.

\vspace{-0.6em}
\subsection{Post-Processing After Beamforming}
After the high-resolution beamforming network, the beamforming branch towards $\theta_q$ is mainly affected by single dominant CFO ${\hat{\varphi}_{q}} ={{\hat{f}}_{d}}\cos {{\theta }_{q}}+\hat{\xi }$. For the signal in the $m$th block of the $q$th branch, the CFO compensation could be performed as
\begin{align}
 {{\mathbf{\hat{z}}}_{m}}\left( {{\theta }_{q}} \right) ={{e}^{-\text{j}2{\rm  \pi} \hat{\varphi}_{q} \frac{\left( m-1 \right)\left( N+{{N}_{\mathrm{cp}}} \right)}{N}}} \operatorname{diag}\left( 1,\ {{e}^{-\text{j}2{\rm \pi} \hat{\varphi}_{q}\frac{1}{N}}},\ \cdots \!,\ {{e}^{-\text{j}2{\rm \pi} \hat{\varphi}_{q}\frac{N-1}{N}}} \right){{\mathbf{z}}_{m}}\left( {{\theta }_{q}} \right).
\end{align}

Then, the equivalent channel in each branch could be regarded as frequency-selective but time-invariant. The adoption of OFDM system again decomposes this frequency-selective channel into parallel flat-fading subcarrier channels. Finally, the transmitted data symbols are readily recovered through MRC among all the beamforming branches.

\section{Proposed joint estimation algorithm for partly calibrated ULA}
From (\ref{ReceivedSignal}), the received pilot signal in the case of partly calibrated ULA could be written as
\begin{align}{\label{ReceivedSignalUncalibrated}}
\mathbf{Y}=\sum\limits_{l=1}^{L}{\sum\limits_{p=1}^{P}{\mathbf{E}\left( {{\varphi }_{l,p}} \right)\mathbf{Bh}\left( {{\theta }_{l,p}} \right) {\boldsymbol{\alpha}}^{T}\left( \boldsymbol{\varepsilon} \right) {\mathbf{V}}^{T}\left( \theta_{l,p} \right) }}+\mathbf{W}.
\end{align}

Imperfect calibration of the uncalibrated ULA destroys the quasi-orthogonality between steering vectors pointing to two distinct directions, since ${{\mathbf{a}}^{H}}\left( {{\theta }_{1}} \right)\mathbf{a}\left( {{\theta }_{2}}, \boldsymbol{\varepsilon}  \right)={{\mathbf{a}}^{H}}\left( {{\theta }_{1}} \right) {\mathbf{V}}\left( \theta_{2} \right) {\boldsymbol{\alpha}}\left( \boldsymbol{\varepsilon} \right) \approx 0$ may not hold anymore for ${{\theta }_{1}}\ne {{\theta }_{2}}$. Therefore, the ULA response vector might fail to eliminate the inter-direction interference and cannot serve as an efficient beamformer. A new algorithm specially designed for partly calibrated ULA is thereby needed.


\vspace{-0.6em}
\subsection{MSE Performance Analysis of the Joint Estimation Algorithm for Fully Calibrated ULA in Partly Calibrated Case}
In this subsection, we will examine the MSE performance loss if we directly apply the joint estimation algorithm developed for fully calibrated ULA to partly calibrated case.
For ease of derivation, we assume that the channel at each delay shares the same uniform incident AoA set as in~\cite{YR_Zheng2003TC}, i.e., the AoA associated to the $p$th subpath of the $l$th delay is $\theta_{l,p} = 2\pi \frac{p}{P}$ for $p=1,2,\cdots\!,P$.
By denoting ${\theta_{p}} = {\theta_{l,p}}, l=1,2,\cdots\!,L$ and ${{\varphi }_{p}}={{f}_{d}}\cos {{\theta }_{p}}+\xi $, the received pilot signal (\ref{ReceivedSignalUncalibrated}) can be re-expressed as
\begin{align}
{{\mathbf{Y}}} = \sum\limits_{l=1}^{L}{\sum\limits_{{p}=1}^{P}{\mathbf{E}\left( {{\varphi }_{l,{p}}} \right){{\mathbf{B}}}\mathbf{h}\left( {{\theta }_{l,{p}}} \right) {\boldsymbol{\alpha}}^{T}\left( \boldsymbol{\varepsilon} \right) {\mathbf{V}}^{T}\left( \theta_{l,p} \right) }} + {{\mathbf{W}}} = \underbrace{\sum\limits_{p=1}^{P}{\mathbf{E}\left( {{\varphi }_{p}} \right)\mathbf{B}{{\mathbf{h}}_{p}}{{\boldsymbol{\alpha }}^{T}}{{\mathbf{V}}^{T}}\left( {{\theta }_{p}} \right)}}_{{{\mathbf{Y}}_{0}}}+\mathbf{W},
\end{align}
where ${{\mathbf{h}}_{p}} = \sum\nolimits_{l=1}^{L}{\mathbf{h}\left( {{\theta }_{l,{p}}} \right)}$ is an $L\times 1$ vector whose $l$th element is $g_{l,p}$ and ${\boldsymbol{\alpha }} \left( \boldsymbol{\varepsilon} \right)$ is simplified as ${\boldsymbol{\alpha }}$ for conciseness. Irrespective of the inter-subarray mismatches, we still adopt the joint estimation algorithm in Section III-B to estimate the CFO. The beamforming direction ${\tilde{\theta}}_{q}$ is chosen from the set ${{\mathcal{D}}_{\text{IFFT}}}$. Define ${\tilde{\varphi}}_{q}={{\tilde{f}}_{d}}\cos {\tilde{\theta}}_{q}+\tilde{\xi }$ and $\tilde{\varphi }={{\tilde{f}}_{d}}\cos \tilde{\theta }+\tilde{\xi }$. With the impact of inter-subarray mismatches, the cost function $g\left( {{{\tilde{f}}}_{d}}, \tilde{\xi } \right)$ in (\ref{OptimPb1}) could be equivalently expressed as
\begin{align}
g\left( {{{\tilde{f}}}_{d}}, \tilde{\xi } \right)= & \sum\limits_{q=1}^{Q} {\left\| \mathbf{P}_{\mathbf{B}}^{\bot }{{\mathbf{E}}^{H}}\left( {\tilde{\varphi }}_{q} \right)\left( {{\mathbf{Y}}_{0}}+\mathbf{W} \right){{\mathbf{a}}^{*}}\left( {\tilde{\theta }}_{q} \right) \right\|_{2}^{2}}, \ {\tilde{\theta}}_{q} \in {{{\mathcal{D}}}_{\text{IFFT}}}, \nonumber \\
\propto & \int_{\rm{\pi}}^{0} \left\| \mathbf{P}_{\mathbf{B}}^{\bot }{{\mathbf{E}}^{H}}\left( {\tilde{\varphi }} \right)\left( {{\mathbf{Y}}_{0}}+\mathbf{W} \right){{\mathbf{a}}^{*}}\left( {\tilde{\theta }} \right) \right\|_{2}^{2} d\cos \tilde{\theta } \nonumber \\
\approx & \underbrace{\int_{0}^{\rm{\pi}}{ {{\mathbf{a}}^{T}}\left( {\tilde{\theta }} \right)\mathbf{Y}_{0}^{H}\mathbf{E}\left( {\tilde{\varphi }} \right)\mathbf{P}_{\mathbf{B}}^{\bot }{{\mathbf{E}}^{H}}\left( {\tilde{\varphi }} \right){{\mathbf{Y}}_{0}}{{\mathbf{a}}^{*}}\left( {\tilde{\theta }} \right) \sin \tilde{\theta }d\tilde{\theta }}}_{{{g}_{0}}} \nonumber \\
+ & \underbrace{2\Re \left\{ \int_{0}^{\rm{\pi}}{ {{\mathbf{a}}^{T}}\left( {\tilde{\theta }} \right)\mathbf{Y}_{0}^{H}\mathbf{E}\left( {\tilde{\varphi }} \right)\mathbf{P}_{\mathbf{B}}^{\bot }{{\mathbf{E}}^{H}}\left( {\tilde{\varphi }} \right)\mathbf{W}{{\mathbf{a}}^{*}}\left( {\tilde{\theta }} \right) \sin \tilde{\theta }d\tilde{\theta }} \right\}}_{{{g}_{\mathrm{n}}}},
\end{align}
where ${{g}_{0}}$ and ${{g}_{\mathrm{n}}}$ represent the contribution of inter-direction interference and that of noise, respectively. Note that all the first-order moments of $\mathbf{W}$ are zero, which leads to $E\left\{ \frac{\partial {{g}_{0}}}{\partial {{{\tilde{f}}}_{d}}}\frac{\partial {{g}_{\mathrm{n}}}}{\partial {{{\tilde{f}}}_{d}}} \right\}=E\left\{ \frac{\partial {{g}_{0}}}{\partial \tilde{\xi }}\frac{\partial {{g}_{\mathrm{n}}}}{\partial \tilde{\xi }} \right\}=E\left\{ \frac{{{\partial }^{2}}{{g}_{\mathrm{n}}}}{\partial {{{\tilde{f}}}_{d}}^{2}} \right\}=E\left\{ \frac{{{\partial }^{2}}{{g}_{\mathrm{n}}}}{\partial {{{\tilde{f}}}_{d}}\partial \tilde{\xi }} \right\}=E\left\{ \frac{{{\partial }^{2}}{{g}_{\mathrm{n}}}}{\partial {{{\tilde{\xi }}}^{2}}} \right\}=0$.
As a result, we derive $E\left\{ {{\left( \frac{\partial g}{\partial {{{\tilde{f}}}_{d}}} \right)}^{2}} \right\} =E\left\{ {{\left( \frac{\partial {{g}_{\mathrm{n}}}}{\partial {{{\tilde{f}}}_{d}}} \right)}^{2}} \right\}+E\left\{ {{\left( \frac{\partial {{g}_{0}}}{\partial {{{\tilde{f}}}_{d}}} \right)}^{2}} \right\} \gtrapprox E\left\{ {{\left( \frac{\partial {{g}_{\mathrm{n}}}}{\partial {{{\tilde{f}}}_{d}}} \right)}^{2}} \right\}+\left( E\left\{ {{\frac{\partial {{g}_{0}}}{\partial {{{\tilde{f}}}_{d}}} }} \right\}\right)^{2}$, $E\left\{ {{\left( \frac{\partial g}{\partial \tilde{\xi }} \right)}^{2}} \right\} =E\left\{ {{\left( \frac{\partial {{g}_{\mathrm{n}}}}{\partial \tilde{\xi }} \right)}^{2}} \right\}+E\left\{ {{\left( \frac{\partial {{g}_{0}}}{\partial \tilde{\xi }} \right)}^{2}} \right\} \gtrapprox E\left\{ {{\left( \frac{\partial {{g}_{\mathrm{n}}}}{\partial {{{\tilde{\xi}}}}} \right)}^{2}} \right\}+\left( E\left\{ {{\frac{\partial {{g}_{0}}}{\partial {{{\tilde{\xi}}}}} }} \right\}\right)^{2}$, $E\left\{ \frac{{{\partial }^{2}}g}{\partial {{{\tilde{f}}}_{d}}^{2}} \right\} =E\left\{ \frac{{{\partial }^{2}}{{g}_{0}}}{\partial {{{\tilde{f}}}_{d}}^{2}} \right\}$, $E\left\{ \frac{{{\partial }^{2}}g}{\partial {{{\tilde{f}}}_{d}}\partial \tilde{\xi }} \right\} =E\left\{ \frac{{{\partial }^{2}}{{g}_{0}}}{\partial {{{\tilde{f}}}_{d}}\partial \tilde{\xi }} \right\}$ and $E\left\{ \frac{{{\partial }^{2}}g}{\partial {{{\tilde{\xi }}}^{2}}} \right\} =E\left\{ \frac{{{\partial }^{2}}{{g}_{0}}}{\partial {{{\tilde{\xi }}}^{2}}} \right\}$.

Denote $\boldsymbol{\tilde{\phi}} \!=\! \left[ {\tilde{f}}_{d}, \tilde{\xi} \right]^{T}$ and $\boldsymbol{\phi} \!=\! \left[ {f}_{d}, \xi \right]^{T}$. Define $a_{11}^{0} \!=\! {E{\left. \left\{ \frac{\partial {{g}_{0}}}{\partial \tilde{f}_d} \right\} \right|}_{ \boldsymbol{\tilde{\phi}} = \boldsymbol{\phi} }}$, $a_{12}^{0} \!=\! {E{\left. \left\{ \frac{\partial {{g}_{0}}}{\partial \tilde{\xi }} \right\} \right|}_{ \boldsymbol{\tilde{\phi}} = \boldsymbol{\phi} }}$, $a_{11}^{\mathrm{n}} \!=\! E{{\left. \left\{ {{\left( \frac{\partial {{g}_{\mathrm{n}}}}{\partial {{{\tilde{f}}}_{d}}} \right)}^{2}} \right\} \right|}_{\boldsymbol{\tilde{\phi}} = \boldsymbol{\phi} }}$, $a_{12}^{\mathrm{n}} \!=\! E{{\left. \left\{ {{\left( \frac{\partial {{g}_{\mathrm{n}}}}{\partial {{{\tilde{\xi}}}}} \right)}^{2}} \right\} \right|}_{\boldsymbol{\tilde{\phi}} = \boldsymbol{\phi} }}$, ${{a}_{21}} \!=\! {{\left. E\left\{ \frac{{{\partial }^{2}}{{g}_{0}}}{\partial \tilde{f}_{d}^{2}} \right\} \right|}_{\boldsymbol{\tilde{\phi}} = \boldsymbol{\phi} }}$, ${{a}_{22}} \!=\! {{\left. E\left\{ \frac{{{\partial }^{2}}{{g}_{0}}}{\partial {{{\tilde{f}}}_{d}}\partial \tilde{\xi }} \right\} \right|}_{\boldsymbol{\tilde{\phi}} = \boldsymbol{\phi} }}$, ${{a}_{23}} \!=\! {{\left. E\left\{ \frac{{{\partial }^{2}}{{g}_{0}}}{\partial {{{\tilde{\xi }}}^{2}}} \right\} \right|}_{\boldsymbol{\tilde{\phi}} = \boldsymbol{\phi} }}$. The detailed derivation for the expression of $a_{11}^{0}, a_{12}^{0}, a_{11}^{\mathrm{n}}, a_{12}^{\mathrm{n}}, a_{21}, a_{22}, a_{23}$ can be found in Appendix~\ref{MSEDerivation}.

With some tedious derivation in Appendix~\ref{CrossTerm}, we can further prove ${{a}_{23}}\approx 2{{a}_{21}}\gg {{a}_{22}}$, which suggests that ${{a}_{22}}$ is negligible and that DFO and OFO estimations are quasi-independent of each other. As a result, the MSE of CFO estimation could be finally expressed as~\cite{W_Zhang2013TSP, G_Wang2011TWC}
\begin{align}
 \text{MSE}\left\{ {{{\tilde{f}}}_{d}} \right\} & ={{\left. \frac{E\left\{ {{\left( \frac{\partial {{g}_{0}}}{\partial {{{\tilde{f}}}_{d}}} \right)}^{2}}+{{\left( \frac{\partial {{g}_{\mathrm{n}}}}{\partial {{{\tilde{f}}}_{d}}} \right)}^{2}} \right\}}{{{\left( E\left\{ \frac{{{\partial }^{2}}{{g}_{0}}}{\partial \tilde{f}_{d}^{2}} \right\} \right)}^{2}}} \right|}_{\boldsymbol{\tilde{\phi}} = \boldsymbol{\phi} }}  \gtrapprox \frac{\left( a_{11}^{0} \right)^{2} + a_{11}^{\mathrm{n}}}{\left( a_{21} \right)^{2}}, \\
 \text{MSE}\left\{ {\tilde{\xi }} \right\} & = {{\left. \frac{E\left\{ {{\left( \frac{\partial {{g}_{0}}}{\partial \tilde{\xi }} \right)}^{2}}+{{\left( \frac{\partial {{g}_{\mathrm{n}}}}{\partial \tilde{\xi }} \right)}^{2}} \right\}}{{{\left( E\left\{ \frac{{{\partial }^{2}}{{g}_{0}}}{\partial {{{\tilde{\xi }}}^{2}}} \right\} \right)}^{2}}} \right|}_{\boldsymbol{\tilde{\phi}} = \boldsymbol{\phi} }} \gtrapprox \frac{\left( a_{12}^{0} \right)^{2} + a_{12}^{\mathrm{n}}}{\left( a_{23} \right)^{2}},
\end{align}
where
\begin{align}
a_{11}^{0} & \approx N \int_{0}^{ \rm{\pi} }\int_{0}^{ \rm{\pi} }\left( \frac{1}{3}-\frac{{{\left( { \rm{\pi} } {{f}_{d}}\tilde{x} \right)}^{2}}}{30} \right) \sin \left( { \rm{\pi} }{{f}_{d}}\tilde{x} \right){{\boldsymbol{\alpha }}^{T}}{{\mathbf{A}}_{b}}{{\boldsymbol{\alpha }}^{\text{*}}}\sin 2\tilde{\theta }d\tilde{\theta }d{{\theta }_{p}}, \\ \label{a110}
a_{12}^{0} & \approx 2N \int_{0}^{ \rm{\pi} }\int_{0}^{ \rm{\pi} }\left( \frac{1}{3}-\frac{{{\left( { \rm{\pi} } {{f}_{d}}\tilde{x} \right)}^{2}}}{30} \right) \sin \left( { \rm{\pi} }{{f}_{d}}\tilde{x} \right){{\boldsymbol{\alpha }}^{T}}{{\mathbf{A}}_{b}}{{\boldsymbol{\alpha }}^{\text{*}}}\sin \tilde{\theta }d\tilde{\theta }d{{\theta }_{p}}, \\ \label{a120}
a_{11}^{\mathrm{n}} & \approx \frac{2{\rm{\pi}}N \sigma_{\mathrm{n}}^{2}} {3\tilde{d}} \int_{0}^{ \rm{\pi} } {\int_{0}^{ \rm{\pi} } {{{\cos }^{2}}\tilde{\theta } {{\boldsymbol{\alpha }}^{T}} {{\mathbf{A}}_{b}}{{\boldsymbol{\alpha }}^{*}}\sin \tilde{\theta } d\tilde{\theta }}d{{\theta }_{p}}}, \\ \label{a11n}
a_{12}^{\mathrm{n}} & \approx \frac{2{\rm{\pi}}N \sigma_{\mathrm{n}}^{2}}{3\tilde{d}} \int_{0}^{ \rm{\pi} }{\int_{0}^{ \rm{\pi} } {{{\boldsymbol{\alpha }}^{T}}{{\mathbf{A}}_{b}} {{\boldsymbol{\alpha }}^{*}}\sin \tilde{\theta } d\tilde{\theta }}d{{\theta }_{p}}}, \\ \label{a12n} 
{{a}_{21}} & \approx \frac{2{\rm{\pi}}N}{3} \int_{0}^{ \rm{\pi} } {\int_{0}^{ \rm{\pi} }{{{\cos }^{2}} \tilde{\theta }{{\boldsymbol{\alpha }}^{T}} {{\mathbf{A}}_{b}} {{\boldsymbol{\alpha }}^{\text{*}}}\sin \tilde{\theta }d\tilde{\theta }} d{{\theta }_{p}}}, \\ \label{a21}
{{a}_{22}}& \approx \frac{2{\rm{\pi}}N}{3} \int_{0}^{ \rm{\pi} } \int_{0}^{ \rm{\pi} }{2\cos \tilde{\theta } {{\boldsymbol{\alpha }}^{T}} {{\mathbf{A}}_{b}} {{\boldsymbol{\alpha }}^{\text{*}}}\sin \tilde{\theta }d\tilde{\theta }} d{{\theta }_{p}}, \\ \label{a22}
{{a}_{23}} & \approx \frac{2{\rm{\pi}}N}{3} \int_{0}^{ \rm{\pi} } \int_{0}^{ \rm{\pi} }{{{\boldsymbol{\alpha }}^{T}} {{\mathbf{A}}_{b}} {{\boldsymbol{\alpha }}^{\text{*}}}\sin \tilde{\theta }d\tilde{\theta }} d{{\theta }_{p}}. 
\end{align}
Here, the $(p,q)$th element of ${\mathbf{A}}_{b} \in \mathbb{C}^{K\times K}$ is $[{\mathbf{A}}_{b}]_{p,q} \!=\! \frac{{{\sin }^{2}}\left( \chi J\tilde{x} \right)}{{{\sin }^{2}}\left( \chi\tilde{x} \right)} {{e}^{\text{j}2\chi J\tilde{x} \left( q-p \right)}}$, with $\tilde{x} \!=\! \cos \tilde{\theta } \!-\! \cos {{\theta }_{p}}$.

Let us further define
\begin{align}\label{MSE0n}
\text{MS}{{\text{E}}_{0}}\left\{ {{{\tilde{f}}}_{d}} \right\}=\frac{\left( a_{11}^{0} \right)^2}{\left( a_{21} \right)^{2}}, \text{MS}{{\text{E}}_{\mathrm{n}}}\left\{ {{{\tilde{f}}}_{d}} \right\}=\frac{a_{11}^{\mathrm{n}}}{\left( a_{21} \right)^{2}}, \text{MS}{{\text{E}}_{0}}\left\{ {{{\tilde{\xi}}}} \right\}=\frac{\left( a_{12}^{0} \right)^2}{\left( a_{23} \right)^{2}}, \text{MS}{{\text{E}}_{\mathrm{n}}}\left\{ {{{\tilde{\xi}}}} \right\}=\frac{a_{12}^{\mathrm{n}}}{\left( a_{23} \right)^{2}}.
\end{align}

Here, ${{\text{MSE}}_{\mathrm{n}}}\left\{ \cdot \right\}$ can be regarded as the contribution of noise to the MSE and decreases as the signal-to-noise ratio (SNR) increases. In contrast, ${{\text{MSE}}_{0}}\left\{ \cdot \right\}$ reflects the influence of inter-direction interference on MSE.
It is independent of the SNR and appears as the MSE floor at high SNRs.
Thus, the latter dominates MSE performance under high SNR region while the former is dominant under low and moderate SNRs. Besides, it must be pointed out that ${{\text{MSE}}_{0}}\left\{ \cdot \right\}$ in (\ref{MSE0n}) is actually a lower bound of the real MSE floor, since we have approximated ${E\left\{ {{\left( \frac{\partial {{g}_{0}}}{\partial {{{\tilde{f}}}_{d}}} \right)}^{2}} \right\}}$ and ${E\left\{ {{\left( \frac{\partial {{g}_{0}}}{\partial {{{\tilde{\xi}}}}} \right)}^{2}} \right\}}$ by $\left( {E\left\{ {\frac{\partial {{g}_{0}}}{\partial {{{\tilde{f}}}_{d}}} } \right\}} \right)^{2}$ and $\left( {E\left\{ {\frac{\partial {{g}_{0}}}{\partial {{{\tilde{\xi}}}}} } \right\}} \right)^{2}$, respectively, for ease of MSE derivation.

Simulations will show that ${{\text{MSE}}_{0}}\left\{ \cdot \right\}$ is evident for large nDFOmax or large number of subarrays, which will cause significant uncompensated residual CFOs and thereby considerably exacerbate the subsequent data detection performance. Thus, it is necessary to amend the current algorithm so that it adapts to partly calibrated ULA. This procedure will be developed in detail in the next subsection.
Simplifying ${{\text{MSE}}_{\mathrm{n}}} \left\{ \cdot \right\}$ and ${{\text{MSE}}_{0}} \left\{ \cdot \right\}$ in (\ref{MSE0n}) into a more concise form is quite an arduous task. Nevertheless, for fully calibrated ULA, we have the following proposition.

\begin{proposition}
In the case of fully calibrated ULA, as the number of antennas $M$ increases, the asymptotic estimation MSEs of both DFO and OFO decrease linearly with the reduction in noise power $\sigma_{\mathrm{n}}^2$, and the asymptotic MSE of DFO is approximately twice that of OFO, i.e.,
\begin{align}\label{MSE_ULA}
 {{\text{MSE}}_{\mathrm{n}}}\left\{ {{{\tilde{f}}}_{d}} \right\} \approx 2{{\text{MSE}}_{\mathrm{n}}}\left\{ {\tilde{\xi }} \right\} \approx \frac{3\sigma _{\mathrm{n}}^{2}}{{{ \rm{\pi} }^{2}}MN}, \ {{\text{MSE}}_{0}}\left\{ {{{\tilde{f}}}_{d}} \right\} \approx 0, \ {{\text{MSE}}_{0}}\left\{ {\tilde{\xi }} \right\} \approx 0.
\end{align}
The detailed derivation could be found in Appendix~\ref{MSE_ULA_proof}. Within expectation, no remarkable MSE floor is observed. Besides, the improvement of SNR condition, enlargement of antenna array or increase of the number of subcarriers help enhance both DFO and OFO estimation performance.
\end{proposition}

\vspace{-0.6em}
\subsection{Joint Estimation Algorithm for Partly Calibrated ULA}
In this subsection, the COBP will be introduced in the design of beamforming network to combat the detrimental effects of imperfect calibration, so that the algorithm could be extended to partly calibrated case.
The diagram of this adapted procedure is illustrated in Fig. 3.
In contrast to Fig. 2, the main difference lies in that the estimation of CFO, COBP and channel is performed prior to the high-resolution beamforming network.

\begin{figure}[t]
\setlength{\abovecaptionskip}{-0.5cm}
\setlength{\belowcaptionskip}{-1cm}
\begin{center}
\includegraphics[width=100mm]{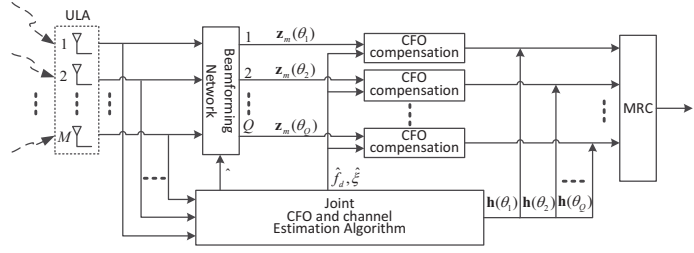}
\end{center}
\caption{ Diagram of the receiver design in the case of partly calibrated ULA.}
\end{figure}

When the ULA is partly calibrated, we adopt the modified beamformer
\begin{align}
\mathbf{b}\left( \theta , \boldsymbol{\varepsilon } \right)= \mathbf{V}\left( \theta  \right)\boldsymbol{\beta }
\end{align}
to perform beamforming. Here, the COBP $\boldsymbol{\beta }$ is introduced to repair the loss of orthogonality caused by inter-subarray mismatches. To some extent, $\boldsymbol{\beta }$ can be regarded as the counterpart of inter-subarray gain and phase mismatches $\boldsymbol{\alpha}(\boldsymbol{\varepsilon}) $.
Let the received signal pass through the modified beamformer $\mathbf{b}\left( \theta , \boldsymbol{\varepsilon } \right)$. Then, the resulting signal $\mathbf{z}\left( \theta  \right) =\mathbf{Y}{{\mathbf{b}}^{\text{*}}}\left( \theta , \boldsymbol{\varepsilon } \right)$ is given by
\begin{align}
\mathbf{z}\left( \theta  \right) & =\underbrace{ \kappa\left( \theta  \right) \sum\limits_{l,p, {{\theta }_{l,p}}=\theta }{\mathbf{E}\left( {{\varphi }_{l,p}} \right)\mathbf{Bh}\left( {{\theta }_{l,p}} \right)}}_{\mathrm{desired \ signal}} +\underbrace{\sum\limits_{l', p', {{\theta }_{l',p'}}\ne \theta }{\mathbf{E}\left( {{\varphi }_{l',p'}} \right)\mathbf{Bh}\left( {{\theta }_{l',p'}} \right){{\mathbf{a}}^{T}}\left( {{\theta }_{l', p'}}, \boldsymbol{\varepsilon } \right){{\mathbf{b}}^{\text{*}}}\left( \theta , {\boldsymbol{\varepsilon}}  \right)}}_{\mathrm{interference}} \nonumber \\
& + \underbrace{\mathbf{W}{{\mathbf{b}}^{\text{*}}}\left( \theta , \boldsymbol{\varepsilon } \right)}_{\mathrm{noise}},
\end{align}
where $\kappa\left( \theta  \right) = {{\mathbf{a}}^{T}}\left( {{\theta }}, \boldsymbol{\varepsilon } \right) {{\mathbf{b}}^{\text{*}}}\left( \theta , {\boldsymbol{\varepsilon}}  \right) = {{\boldsymbol{\alpha}}^{T}}\left( \boldsymbol{\varepsilon } \right) {{\mathbf{V}}^{T}}\left( {{\theta }} \right) \mathbf{V}\left( \theta  \right)\boldsymbol{\beta }$. Besides, the first term is the desired signal from direction $\theta$, while the second and third terms represent the inter-direction interference and noise after beamforming respectively. Since the rectified beamformer is expected to combat array mismatches, there should be ${{\mathbf{a}}^{T}}\left( {{\theta }_{l', p'}}, \boldsymbol{\varepsilon } \right){{\mathbf{b}}^{\text{*}}}\left( \theta , {\boldsymbol{\varepsilon}}  \right) \!=\! {{\boldsymbol{\alpha}}^{T}}\left( \boldsymbol{\varepsilon } \right) {{\mathbf{V}}^{T}}\left( {{\theta }_{l', p'}} \right) \mathbf{V}\left( \theta  \right) \boldsymbol{\beta } \approx\! 0$ for ${{\theta }_{l', p'}} \\ \ne \theta$. As a result, the interference from other directions will be prominently mitigated. By ignoring the interference, we arrive at
\begin{align}
\mathbf{z}\left( \theta  \right)=\mathbf{Y}{{\mathbf{V}}^{\text{*}}}\left( \theta  \right){{\boldsymbol{\beta }}^{*}}=\mathbf{E}\left( {{f}_{d}}\cos \theta +\xi  \right)\mathbf{B}\mathbbm{h}\left( \theta  \right)+\mathbf{\tilde{w}}\left( \theta \right),
\end{align}
where $\mathbbm{h}\left( \theta  \right)=\kappa\left( \theta  \right) \sum\nolimits_{{{\theta }_{l,p}}=\theta }{\mathbf{h}\left( {{\theta }_{l,p}} \right)}$
and $\mathbf{\tilde{w}}\left( \theta \right) =\mathbf{W}{{\mathbf{V}}^{\text{*}}} \left( \theta \right) {{\boldsymbol{\beta }}^{*}}$.

Similar to Section III-B, the maximum DFO, OFO and COBP could be found by solving the following minimization problem
\begin{align} \label{OptimBeta}
\left\{{\hat{f}}_{d}, \hat{\xi }, \boldsymbol{\hat{\beta }}\right\} = \arg \underset{\left\{{{\tilde{f}}}_{d}, \tilde{\xi }, \boldsymbol{\tilde{\beta }}\right\}}{\mathop{\min }}\,{{{\boldsymbol{\tilde{\beta }}}}^{H}}{\mathbf{C}\left( {{{\tilde{f}}}_{d}}, \tilde{\xi } \right)}\boldsymbol{\tilde{\beta }},\ s.t.\ \left\| {\boldsymbol{\tilde{\beta }}} \right\|_{2}^{2}=1,
\end{align}
where
\begin{align}
& \mathbf{C}\left( {{{\tilde{f}}}_{d}}, \tilde{\xi } \right) = \sum\limits_{q=1}^{Q}{{\mathbf{V}}^{H}}\left( {{\theta }_{q}} \right){{\mathbf{Y}}^{T}}{{\mathbf{E}}^{H}}\left( \tilde{\varphi}_{q} \right) {{\mathbf{P}_{\mathbf{B}}^{\bot } }^{T}}\mathbf{E}\left( \tilde{\varphi}_{q} \right){{\mathbf{Y}}^{*}}\mathbf{V}\left( {{\theta }_{q}} \right).
\end{align}

The constraint $\big\| {\boldsymbol{\tilde{\beta }}} \big\|_{2}^{2}=1$ is added because otherwise (\ref{OptimBeta}) achieves its minimum at $\boldsymbol{\hat{\beta }}=\mathbf{0}$, which is undesired for the subsequent processing. Moreover, the equivalent channel for the $q$th beamforming branch could be estimated by
\begin{align}\label{MLChannel}
\mathbbm{\hat{h}}\left( \theta_{q} \right)={{\mathbf{B}}^{\dagger }}{{\mathbf{E}}^{H}}\left( {\hat{f}_d} \cos \theta_{q} + \hat{\xi}  \right)\mathbf{Y}{{\mathbf{V}}^{\text{*}}}\left( \theta_{q} \right){{\boldsymbol{\hat{\beta} }}^{\text{*}}}.
\end{align}

For a given trial value pair ${{{\tilde{f}}}_{d}}$ and $\tilde{\xi}$, (\ref{OptimBeta}) is equivalent to minimizing $H\left( \boldsymbol{\tilde{\beta} } \right)= {{{\boldsymbol{\tilde{\beta }}}}^{H}}{\mathbf{C}\left( {{{\tilde{f}}}_{d}}, \tilde{\xi } \right)}\boldsymbol{\tilde{\beta }} \\
+\mu \left( 1-{{\boldsymbol{\tilde{\beta} }}^{H}}\boldsymbol{\tilde{\beta} } \right)$, where $\mu$ is the Lagrange Multiplier. By means of the first-order condition, the optimal solution of $\boldsymbol{\tilde{\beta} }$ is given by $\boldsymbol{\hat{\beta}}={{\mathbf{v}}_{\min }}\left( {\mathbf{C}\left( {{{\tilde{f}}}_{d}}, \tilde{\xi } \right)}\right)$ and the corresponding minimum attained at $\boldsymbol{\hat{\beta}}$ is ${{\left. H\left( {\boldsymbol{\tilde{\beta }}} \right) \right|}_{\boldsymbol{\tilde{\beta }}=\boldsymbol{\hat{\beta }}}}={{\boldsymbol{\hat{\beta }}}^{H}}\mathbf{C}\left( {{{\tilde{f}}}_{d}},\tilde{\xi } \right)\boldsymbol{\hat{\beta }}={{\lambda }_{\min }}\left( \mathbf{C}\left( {{{\tilde{f}}}_{d}},\tilde{\xi } \right) \right)$.

Therefore, (\ref{OptimBeta}) could be further decomposed into
\begin{eqnarray}
\left\{
\begin{array}{lll}
\!\! \left\{{{\hat{f}}}_{d}, \hat{\xi }\right\}& \!\!\!\!=\arg \underset{\left\{{{\tilde{f}}}_{d}, \tilde{\xi}\right\}}{\mathop{\min }}\,{{\lambda }_{\min }}\left( {\mathbf{C}\left( {{{\tilde{f}}}_{d}}, \tilde{\xi } \right)} \right), \\
\!\! \ \boldsymbol{\hat{\beta }}&\!\!\!\!={{\mathbf{v}}_{\min }}\left( {\mathbf{C}\left( {{{\hat{f}}}_{d}}, \hat{\xi } \right)} \right).
\end{array}
\right.
\end{eqnarray}

Note that although the algorithm in Section III is designed for fully calibrated ULA, it could also provide valid coarse CFO estimates $\left[ {{f}_{dc}}, \ {{\xi }_{c}} \right]$ in the presence of inter-subarray mismatches. Based on this coarse estimation result, one-tap adjustment via Taylor series expansion is sufficient to obtain fine DFO and OFO estimates. Specifically, denote $\mathbf{\tilde{c}}\left( {{\theta }_{q}} \right)=\mathbf{E}\left( {\varphi}_{qc} \right){{\mathbf{Y}}^{*}}\mathbf{V}\left( {{\theta }_{q}} \right)$, where ${\varphi}_{qc} = f_{dc} \cos{{\theta }_{q}} + \xi_c$, and define $\Delta {\tilde{\varphi}_{q}} = \Delta \tilde f_{d} \cos{{\theta }_{q}} + \Delta \tilde\xi$, where $\Delta \tilde f_{d} = \tilde{f_d} - f_{dc}$ and $\Delta \tilde\xi = \tilde{\xi} - \xi_c$ represent the trial residual nDFOmax and trial residual nOFO, respectively. By substituting $\mathbf{E }\left( {{\tilde{\varphi}}}_{q} \right)$ with its Taylor series expansion at $\left[ {{f}_{dc}}, \ {{\xi }_{c}} \right]$, $\mathbf{C}\left( {{{\tilde{f}}}_{d}}, \tilde{\xi } \right)$ could be approximated as
\begin{align}
\mathbf{C} \left( {{{\tilde{f}}}_{d}}, \tilde{\xi } \right) \approx \underbrace{ \sum\limits_{q=1}^{Q}{ {{\mathbf{T}}_{0}}\left( {{\theta }_{q}} \right) }}_{\mathbf{\Upsilon}} + \underbrace{\sum\limits_{q=1}^{Q}{ \Delta {\tilde{\varphi}_{q}}  {{\mathbf{T}}_{1}}\left( {{\theta }_{q}} \right)+{{\left( \Delta {\tilde{\varphi}_{q}} \right)}^{2}}{{\mathbf{T}}_{2}}\left( {{\theta }_{q}} \right) }}_{\mathbf{\Xi}},
\end{align}
where
\begin{align*}
  {{\mathbf{T}}_{0}}\left( {{\theta }_{q}} \right) &={{{\mathbf{\tilde{c}}}}^{H}}\left( {{\theta }_{q}} \right) {\mathbf{P}_{\mathbf{B}}^{\bot }}^{T}  \mathbf{\tilde{c}}\left( {{\theta }_{q}} \right), \
{{\mathbf{T}}_{1}}\left( {{\theta }_{q}} \right) ={{{\mathbf{\tilde{c}}}}^{H}}\left( {{\theta }_{q}} \right) \left( {{\mathbf{D}}^{H}}{{\mathbf{P}_{\mathbf{B}}^{\bot } }^{T}} + {{ \mathbf{P}_{\mathbf{B}}^{\bot } }^{T}}\mathbf{D} \right) \mathbf{\tilde{c}}\left( {{\theta }_{q}} \right), \\
 {{\mathbf{T}}_{2}}\left( {{\theta }_{q}} \right)&= {{{\mathbf{\tilde{c}}}}^{H}}\left( {{\theta }_{q}} \right) \left( {{\mathbf{D}}^{H}}{{ \mathbf{P}_{\mathbf{B}}^{\bot } }^{T}}\mathbf{D} + \frac{ {{\mathbf{D}}^{2H}}{{ \mathbf{P}_{\mathbf{B}}^{\bot }}^{T}} + {{ \mathbf{P}_{\mathbf{B}}^{\bot } }^{T}}{{\mathbf{D}}^{2}}}{2} \right) \mathbf{\tilde{c}}\left( {{\theta }_{q}} \right).
\end{align*}

Let $\mathcal{A}$ and $\mathcal{B}$ be two arbitrary full-rank matrices,
and denote $\epsilon$ as a sufficiently small perturbation term. Then, according to the perturbation theory~\cite{bJH_Wilkinson1965, W_Zhang2016TSP}, there holds ${{\lambda }_{\min }}\left( \mathcal{A} \!+\! \epsilon\mathcal{B} \right) \!\approx\! {{\lambda }_{\min }}\left( \mathcal{A} \right) \!+\! \epsilon {{\mathbf{v}}_{\mathrm{min}}^{H}}\left( \mathcal{A} \right) \mathcal{B} {{\mathbf{v}}_{\mathrm{min}}}\left( \mathcal{A} \right)$. Therefore, denoting $\mathbf{v} \!=\! {\mathbf{v }_{\min }}\left( {\mathbf{\Upsilon}} \right)$, we have
\begin{align}
 {{\lambda }_{\min }}\left( \mathbf{C}\left( {{{\tilde{f}}}_{d}}, \tilde{\xi } \right) \right) & \approx {{\lambda }_{\min }}\left( {\mathbf{\Upsilon}} + {\mathbf{\Xi}} \right) \approx {{\lambda }_{\min }}\left( {\mathbf{\Upsilon}} \right) +{{\mathbf{v}}^{H}} {\mathbf{\Xi}} \mathbf{v} \nonumber \\
 & = {{t}_{0}}+{{t}_{11}}\Delta {\tilde{f}_{d}}+{{t}_{12}}\Delta \tilde\xi+{{t}_{21}}{{\left( \Delta {\tilde{f}_{d}} \right)}^{2}} +{{t}_{22}}\Delta {\tilde{f}_{d}}\Delta \tilde\xi+{{t}_{23}}{{\left( \Delta \tilde\xi  \right)}^{2}},
\end{align}
where
\begin{align*}
{{t}_{0}} & ={{\lambda }_{\min }} \left( {\mathbf{\Upsilon}} \right), \ \ {{t}_{11}} = {{\mathbf{v}}^{H}} \sum\nolimits_{q=1}^{Q}{\cos{{\theta }_{q}}{{\mathbf{T}}_{1}}\left( {{\theta }_{q}} \right)}\mathbf{v}, \ \ {{t}_{12}} ={{\mathbf{v}}^{H}} \sum\nolimits_{q=1}^{Q}{{{\mathbf{T}}_{1}}\left( {{\theta }_{q}} \right)} \mathbf{v}, \nonumber \\
{{t}_{21}} & ={{\mathbf{v}}^{H}} \sum\nolimits_{q=1}^{Q}{{{\cos}^{2}}{{\theta }_{q}}{{\mathbf{T}}_{2}}\left( {{\theta }_{q}} \right)} \mathbf{v}, \ \
 {{t}_{22}} ={{\mathbf{v}}^{H}} \sum\nolimits_{q=1}^{Q}{2\cos{{\theta }_{q}}{{\mathbf{T}}_{2}}\left( {{\theta }_{q}} \right)} \mathbf{v}, \ \ {{t}_{23}} ={{\mathbf{v}}^{H}} \sum\nolimits_{q=1}^{Q}{{{\mathbf{T}}_{2}}\left( {{\theta }_{q}} \right)}\mathbf{v}.
\end{align*}

Similar to (\ref{QuadraticFun1}), the optimal residual nDFOmax $\Delta {{f}_{d}}$ and residual nOFO $\Delta \xi $ are given by
\begin{align}\label{CFOSolution2}
\left[ \begin{matrix}
   \Delta \hat{f}_{d}  \\
   \Delta \hat{\xi}  \\
\end{matrix} \right]=-{{\left[ \begin{matrix}
   2{{t}_{21}} & {{t}_{22}}  \\
   {{t}_{22}} & 2{{t}_{23}}  \\
\end{matrix} \right]}^{-1}}\left[ \begin{matrix}
   {{t}_{11}}  \\
   {{t}_{12}}  \\
\end{matrix} \right],
\end{align}
and the final CFO estimates could be calculated as ${{\hat{f}}_{d}}={{f}_{dc}}+\Delta {\hat{f}_{d}}$ and $\hat{\xi }={{\xi }_{c}}+\Delta \hat{\xi} $.

In summary, the whole estimation process with partly calibrated massive ULA can be described as follows.
\begin{itemize}
  \item \emph{Step-1, coarse CFO estimation}: We first perform beamforming irrespective of inter-subarray mismatches, and get the coarse estimates ${f}_{dc}$ and ${\xi}_{c}$ with the algorithm in Section III-B.
  \item \emph{Step-2, one-tap adjustment via Taylor series expansion}: The inter-subarray gain and phase mismatches are taken into account and the joint estimation algorithm in section IV-B is used to jointly estimate CFO and COBP. The fine CFO estimates can be obtained from (\ref{CFOSolution2}) via two-dimensional Taylor series expansion by setting $\left[ {f}_{dc}, {\xi}_{c} \right]$ as an expansion point.
  \item \emph{Step-3, calculation of COBP}: Once the estimated nDFOmax ${{\hat{f}}_{d}}$ and nOFO $\hat{\xi }$ are obtained, the COBP can be directly calculated as $\boldsymbol{\hat{\beta }}={{\mathbf{v}}_{\min }}\left( \mathbf{C}\left( {{{\hat{f}}}_{d}}, \hat{\xi } \right) \right)$.
  \item \emph{Step-4, computation of the equivalent channel}: Based on the estimates of CFO and COBP, the equivalent channel for the $q$th beamforming branch is readily computed by (\ref{MLChannel}).
\end{itemize}

\section{Derivation of the Cram\'{e}r-Rao Bound}
In this section, we will derive the CRB of CFO estimation. The derivation will be carried out in the case of partly calibrated ULA, but the result also applies to the fully calibrated ULA, wherein the inter-subarray gain and phase mismatches vector $\boldsymbol{\alpha} \left( \boldsymbol{\varepsilon} \right)$ is reduced to scalar $1$.

We first reformulate (\ref{ReceivedSignalUncalibrated}) as
\begin{align}
\mathbf{Y} =\sum\limits_{l=1}^{L}{\sum\limits_{p=1}^{P}{\mathbf{E}\left( {{\varphi }_{l, p}} \right)\mathbf{BH}\left( {{\theta }_{l, p}} \right)\mathbf{G}}}+\mathbf{W},
\end{align}
where $\mathbf{H}\left( {{\theta }_{l, p}} \right) = \mathbf{h}\left( {{\theta }_{l,p}} \right){{\mathbf{a}}^{T}}\left( {{\theta }_{l, p}} \right)$ and $\mathbf{G} =\operatorname{diag}\left( \boldsymbol{\alpha }\left( \boldsymbol{\varepsilon } \right) \otimes {{\mathbf{1}}_{J\times 1}} \right)$ such that $\mathbf{a}\left( {{\theta }_{l, p}}, \boldsymbol{\varepsilon } \right) =\mathbf{G}\mathbf{a}\left( {{\theta }_{l, p}} \right)$.

The vectorization of $\mathbf{Y}$ is given by
\begin{align}
\text{vec} \left( \mathbf{Y} \right)  =\sum\limits_{l=1}^{L}{\sum\limits_{p=1}^{P}{\left( {{\mathbf{I}}_{M}}\otimes \mathbf{E}\left( {{\varphi }_{l,p}} \right) \right)\mathbb{C}\left( {{\mathbf{I}}_{L}}\otimes \mathbf{G} \right)\operatorname{vec}\left( {{\mathbf{H}}^{T}}\left( {{\theta }_{l,p}} \right) \right)}} +\text{vec}\left( \mathbf{W} \right),
\end{align}
where $\mathbb{C}=\left[ \begin{matrix}
   {{\mathbf{I}}_{M}}\otimes {{\mathbf{b}}_{1}}, & {{\mathbf{I}}_{M}}\otimes {{\mathbf{b}}_{2}}, & \cdots\!,  & {{\mathbf{I}}_{M}}\otimes {{\mathbf{b}}_{L}}  \\
\end{matrix} \right]$ and $\mathbf{B}=\left[ \begin{matrix}
   {{\mathbf{b}}_{1}}, & {{\mathbf{b}}_{2}}, & \cdots\!,  & {{\mathbf{b}}_{L}}  \\
\end{matrix} \right]$.

We further obtain $E \left\{ \operatorname{vec}\left( {{\mathbf{H}}^{T}}\left( {{\theta }_{l,p}} \right) \right)\operatorname{vec}{{\left( {{\mathbf{H}}^{T}}\left( {{\theta }_{l',p'}} \right) \right)}^{H}} \right\}
={{\delta }_{l-l', \!\ p-p'}}\frac{\sigma_{l}^{2} }{P}{{\mathbf{E}}_{L}^{l}}\otimes \mathbf{R}\left( {{\theta }_{l,p}} \right)$, where ${{\mathbf{E}}_{L}^{l}}$ is a diagonal matrix whose $l$th element is 1 and 0 elsewhere, and $\mathbf{R}\left( {{\theta }_{l,p}} \right)=\mathbf{a}\left( {{\theta }_{l,p}} \right){{\mathbf{a}}^{H}}\left( {{\theta }_{l,p}} \right)$.

Define $\mathbf{U}=\operatorname{blkdiag}\left( {{\mathbf{1}}_{J\times1}}, {{\mathbf{1}}_{J\times1}}, \cdots\!, {{\mathbf{1}}_{J\times1}} \right)\in {{\mathbb{C}}^{M\times K}}$ such that $\mathbf{G} = \operatorname{diag} \left({\mathbf{U}}{\boldsymbol\alpha}\right)$. Define DFO phase rotation vector $\mathbf{e}\left( {{f}_{l,p}} \right) =\left[ 1,\ {{e}^{\text{j}2{\rm\pi} {{f}_{l,p}}\frac{1}{N}}},\ \cdots\!, \ {{e}^{\text{j}2{\rm\pi} {{f }_{l,p}}\frac{N-1}{N}}} \right]^{T}$ and OFO phase rotation vector $\mathbf{e}\left( {\xi} \right) =\left[ 1,\ {{e}^{\text{j}2{\rm\pi} {\xi}\frac{1}{N}}},\ \cdots\!, \ {{e}^{\text{j}2{\rm\pi} {\xi}\frac{N-1}{N}}} \right]^{T}$ such that $\mathbf{E}\left( {{\varphi }_{l,p}} \right) = \operatorname{diag}\left( \mathbf{e}\left( {{f}_{l,p}} \right) \odot {\mathbf{e}}\left( {\xi} \right) \right)$.
Then, there is
\begin{align}{\label{R}}
 \mathbb{R}= & E\left\{ \operatorname{vec}\left( \mathbf{Y} \right)\operatorname{vec}{{\left( \mathbf{Y} \right)}^{H}} \right\} \nonumber \\
 =&\sum\limits_{l=1}^{L} {\sum\limits_{p=1}^{P}{ \frac{{\sigma_{l}^{2}} }{P} \left( {{\mathbf{I}}_{M}}\otimes \mathbf{E}\left( {{\varphi }_{l,p}} \right) \right)\left( {{\mathbf{I}}_{M}}\otimes {{\mathbf{b}}_{{l}}} \right)\mathbf{G} \mathbf{R}\left( {{\theta }_{l,p}} \right){{\mathbf{G}}^{H}}}} {{\left( {{\mathbf{I}}_{M}}\otimes {{\mathbf{b}}_{{l}}} \right)}^{H}}{{\left( {{\mathbf{I}}_{M}}\otimes \mathbf{E}\left( {{\varphi }_{l,p}} \right) \right)}^{H}}+{{\sigma }_{\mathrm{n}}^{2}}{{\mathbf{I}}_{MN}} \nonumber \\
 =& \sum\limits_{l=1}^{L}{\sum\limits_{p=1}^{P}{ \frac{{\sigma_{l}^{2}} }{P} \left( \mathbf{GR}\left( {{\theta }_{l,p}} \right){{\mathbf{G}}^{H}} \right)\otimes \left( \mathbf{E}\left( {{\varphi }_{l,p}} \right){{\mathbf{b}}_{{l}}}\mathbf{b}_{{l}}^{H}{{\mathbf{E}}^{H}}\left( {{\varphi }_{l,p}} \right) \right)}} +{{\sigma }_{\mathrm{n}}^{2}}{{\mathbf{I}}_{MN}} \nonumber \\
 =& \frac{1}{P}\sum\limits_{l=1}^{L}{\sum\limits_{p=1}^{P}{{{\mathbf{R}}_{1,l,p}}\odot {{\mathbf{R}}_{2}}\odot {{\mathbf{R}}_{3}}\odot {{\mathbf{R}}_{4,l}}}}+{{\sigma }_{\mathrm{n}}^{2}}{{\mathbf{I}}_{MN}},
\end{align}
where ${{\mathbf{R}}_{1,l,p}}=\mathbf{R}\left( {{\theta }_{l,p}} \right)\otimes \left( \mathbf{e}\left( {{f}_{l,p}} \right){{\mathbf{e}}^{H}}\left( {{f}_{l,p}} \right) \right)$, ${{\mathbf{R}}_{2}}={{\mathbf{1}}_{M}}\otimes \left( \mathbf{e}\left( \xi  \right){{\mathbf{e}}^{H}}\left( \xi  \right) \right)$, ${{\mathbf{R}}_{3}}=\left( \mathbf{U}{\boldsymbol \alpha }{{\boldsymbol{\alpha }}^{H}}{{\mathbf{U}}^{T}} \right)\otimes {{\mathbf{1}}_{N}}$ and ${{\mathbf{R}}_{4,l}}={{\mathbf{1}}_{M}}\otimes \left( \sigma_{l}^{2} {{\mathbf{b}}_{l}}\mathbf{b}_{l}^{H} \right)$.

Clearly, ${{\mathbf{R}}_{1,l,p}}$ is related to the incident angle ${{\theta }_{l,p}}$ and nDFOmax, ${{\mathbf{R}}_{2}}$ is determined by nOFO, ${{\mathbf{R}}_{3}}$ depends on the inter-subarray gain and phase mismatches, and ${{\mathbf{R}}_{4,l}}$ is deterministic since the training sequence is assumed known at the receiver.
As ${{\theta }_{l,p}}$ follows uniform distribution in $(0, 2\rm{\pi})$, the expectation of ${{\mathbf{R}}_{1,l,p}}$ with respect to ${\theta }_{l,p}$ can be expressed as
\begin{align}{\label{R1}}
{{\mathbf{\tilde{R}}}_{1}}=E\left\{ {{\mathbf{R}}_{1,l,p}} \right\}={{J}_{0}}\left( \mathbf{U}\left( {{f}_{d}} \right) \right),
\end{align}
where ${{J}_{0}}\left(\cdot \right)$ denotes zero-order Bessel function and the $n$th order Bessel function is given by
\begin{align}
{{J}_{n}}\left( x \right)=\frac{1}{2{\rm\pi}}\int_{-{\rm\pi}}^{{\rm\pi}}{\cos \left( x\sin \theta -n\theta  \right)d\theta }.
\end{align}
Detailed derivation of (\ref{R1}) along with the definition of $\mathbf{U}\left( {{f}_{d}} \right)$ could be found in Appendix~\ref{Expectation}.

Substituting ${{\mathbf{R}}_{1,l,p}}$ with ${{\mathbf{\tilde{R}}}_{1}}$, we can simplify (\ref{R}) into
\begin{align}
 \mathbb{R} ={{{\mathbf{\tilde{R}}}}_{1}}\odot {{\mathbf{R}}_{2}}\odot {{\mathbf{R}}_{3}}\odot {{{\mathbf{\tilde{R}}}}_{4}}+{{\sigma }_{\mathrm{n}}^{2}} {{\mathbf{I}}_{MN}},
\end{align}
where ${{\mathbf{\tilde{R}}}_{4}} = \sum\limits_{l=1}^{L}{{{\mathbf{R}}_{4,l}}} ={{\mathbf{1}}_{M}}\otimes \left( \mathbf{B}\mathbf{\Lambda}{{\mathbf{B}}^{H}} \right) \overset{\sigma_{l}^{2} = \frac{1}{L}}{\mathop{=}}\, \frac{1}{L}{{\mathbf{1}}_{M}}\otimes \left( \mathbf{B}{{\mathbf{B}}^{H}} \right)$.

The unknown parameters to be estimated can be listed as $\boldsymbol{\eta }=\left\{ {{f}_{d}}, \xi, \Re \left( {\boldsymbol{\alpha }}^T \right), \Im \left( {\boldsymbol{\alpha }}^T \right), {{\sigma }_{\mathrm{n}}^{2}} \right\}^T$. According to~\cite{P_Stoica1990TASSP, R_Cao2016TSP}, the CRB can be derived from
\begin{align}{\label{CRB}}
{{\left[ \mathbf{CR}{{\mathbf{B}}^{-1}}\left( \boldsymbol{\eta } \right) \right]}_{kl}}=\operatorname{tr}\left[ {{\mathbb{R}}^{-1}}\frac{\partial \mathbb{R}}{\partial {{\eta }_{k}}}{{\mathbb{R}}^{-1}}\frac{\partial \mathbb{R}}{\partial {{ \eta }_{l}}} \right].
\end{align}

The detailed derivation of all the first-order derivatives $\frac{\partial \mathbb{R}}{\partial {{\eta }_{k}}}$ in (\ref{CRB}) can be found in Appendix~\ref{Derivative}. Note that in the case of fully calibrated ULA, $\boldsymbol{\alpha}$ is reduced to scalar $1$, which leads to $\mathbf{R}_{3} \!=\! \mathbf{I}_{MN}$ and $\mathbb{R} = {{{\mathbf{\tilde{R}}}}_{1}}\odot {{\mathbf{R}}_{2}}\odot {{{\mathbf{\tilde{R}}}}_{4}}+{{\sigma }_{n}^{2}}{{\mathbf{I}}_{MN}}$. Moreover, the parameters to be estimated reduce to $\boldsymbol{\eta } = \left\{ {{f}_{d}}, \xi, {{\sigma }_{\mathrm{n}}^{2}} \right\}^T$ and the derivatives of $\mathbf{R}_{3}$ relative to $\boldsymbol{\alpha}$ in (\ref{CRB}) should also be accordingly removed to compute the CRB.

\begin{remark}
It can be seen from (\ref{CRB}) that the CRB acquired at each simulation depends on the realization of the random parameters $\boldsymbol{\eta }$. The CRB results obtained via (\ref{CRB}) under different CFOs and inter-subarray gain and phase mismatches are further averaged numerically to yield the final CRB provided in simulation.
\end{remark}

\section{Simulation Results}
In this section, we will evaluate the performance of our proposed joint estimation algorithms. The terminal HST employs a partly calibrated ULA composed of $M=64$ receive antennas. Unless otherwise stated, the antenna spacing is taken as $d=0.45\lambda$, i.e. $\tilde{d}=\frac{d}{\lambda}=0.45$. Moreover, we assume that the inter-subarray gain mismatch $|\alpha_k|$ follows i.i.d. uniform distribution~\cite{CMS_See2004TSP, Y_GE2017SPAWC} $U\left(\sqrt{1-\sigma_{\alpha}^2}-\sqrt{3}\sigma_{\alpha}, \sqrt{1-\sigma_{\alpha}^2}+\sqrt{3} \sigma_{\alpha}\right)$, where $\sigma_{\alpha}$ stands for the standard deviation of $|\alpha_k|$. In this way, the average array gain is normalized, i.e., $E\big\{|\alpha_k|^2\big\}\!=\!1$. For simulation, we set $|\alpha_k| \!\sim\! U\left(0.8, 1.1875 \right)$.
The total number of subcarriers is taken as $N\!=\!64$, and the first block of each frame serves as pilot, while the rest three blocks are reserved for data transmission. Both the training and data symbols are randomly drawn from 16-QAM constellations. The lengths of channel and CP are set as $L\!=\!8$ and $N_{\mathrm{cp}}\!=\!16$, respectively. For simplicity, the uniform channel PDP, i.e., $\sigma_{l}^{2} \!=\! \frac{1}{L}, l \!=\! 1,2,\cdots \!, L$ is adopted in simulation, yet it should be pointed out that the algorithms do not rely on any specific channel PDP. In fact, we obtained essentially the same performance results for the channels with exponential PDP and the plots are omitted due to the space limitation.
The carrier frequency is fixed as $f_c \!=\! 9\rm{GHz}$, while the block duration is taken as $T_b\!=\!0.1\rm{ms}$. Unless otherwise stated, the HST velocity is assumed to be $480\rm{km/h}$, which translates to $f_d \!=\! 0.4$. The nOFO is randomly generated from $-0.1$ to $0.1$. The beamforming direction $\theta_q$ is drawn from ${\mathcal{D}}_{{\rm{IFFT}}}$.

The MSE for CFO estimation and the symbol error rate (SER) of the recovered data symbols are adopted as the performance metrics. The joint estimation algorithm in Section III for fully calibrated ULA and that in Section IV-B for partly calibrated ULA are referred to as `No-COBP' and `Optimal-COBP', respectively. In the following SER figures, the ideal case with accurate nDFOmax and nOFO knowledge at the receiver will be included as the benchmark.

\begin{figure}[t]
\setlength{\abovecaptionskip}{-0.5cm}
\setlength{\belowcaptionskip}{-1.2cm}
\begin{center}
\includegraphics[width=121mm]{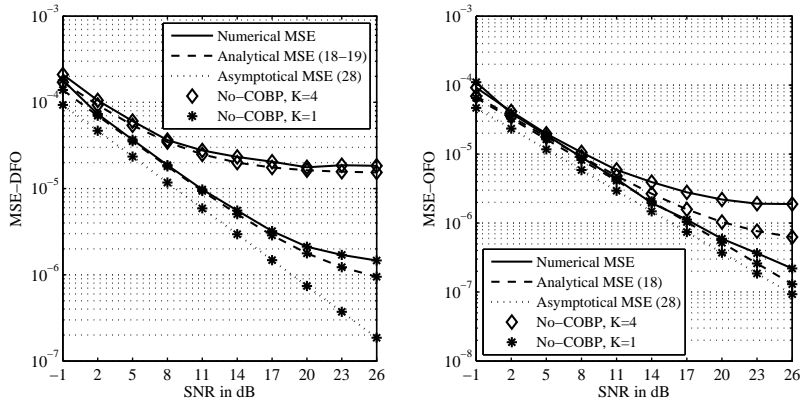}
\end{center}
\caption{ Numerical and analytical MSE comparison of `No-COBP' with fully calibrated ULA ($K=1$) and partly calibrated ULA ($K=4$) at $f_d=0.1$. }
\end{figure}

In Fig. 4, both the numerical, analytical and asymptotical MSEs of `No-COBP' are depicted for fully ($K\!=\!1$) and partly ($K\!=\!4$) calibrated ULA. The nDFOmax is taken as $f_d \!=\! 0.1$. Although the analytical MSE floor is a lower bound of its numerical counterpart, the analytical MSE still well approximates numerical MSE for a wide range of SNR in this example. Meanwhile, we observe an obvious MSE floor at $K\!=\!4$, especially for DFO estimation, which confirms that it is unsuitable to directly apply `No-COBP' to the partly calibrated case and that the new algorithm `Optimal-COBP' is needed. Moreover, a discrepancy between the asymptotical and numerical MSEs exists in this example, which would be reduced by increasing the number of antennas $M$.

\begin{figure}[htbp]
\vspace{-1.5em}
\setlength{\abovecaptionskip}{-0.2cm}
\setlength{\belowcaptionskip}{-0.85cm}
  \centering
  \begin{minipage}{80mm}
  \centering
    \includegraphics[width=70mm]{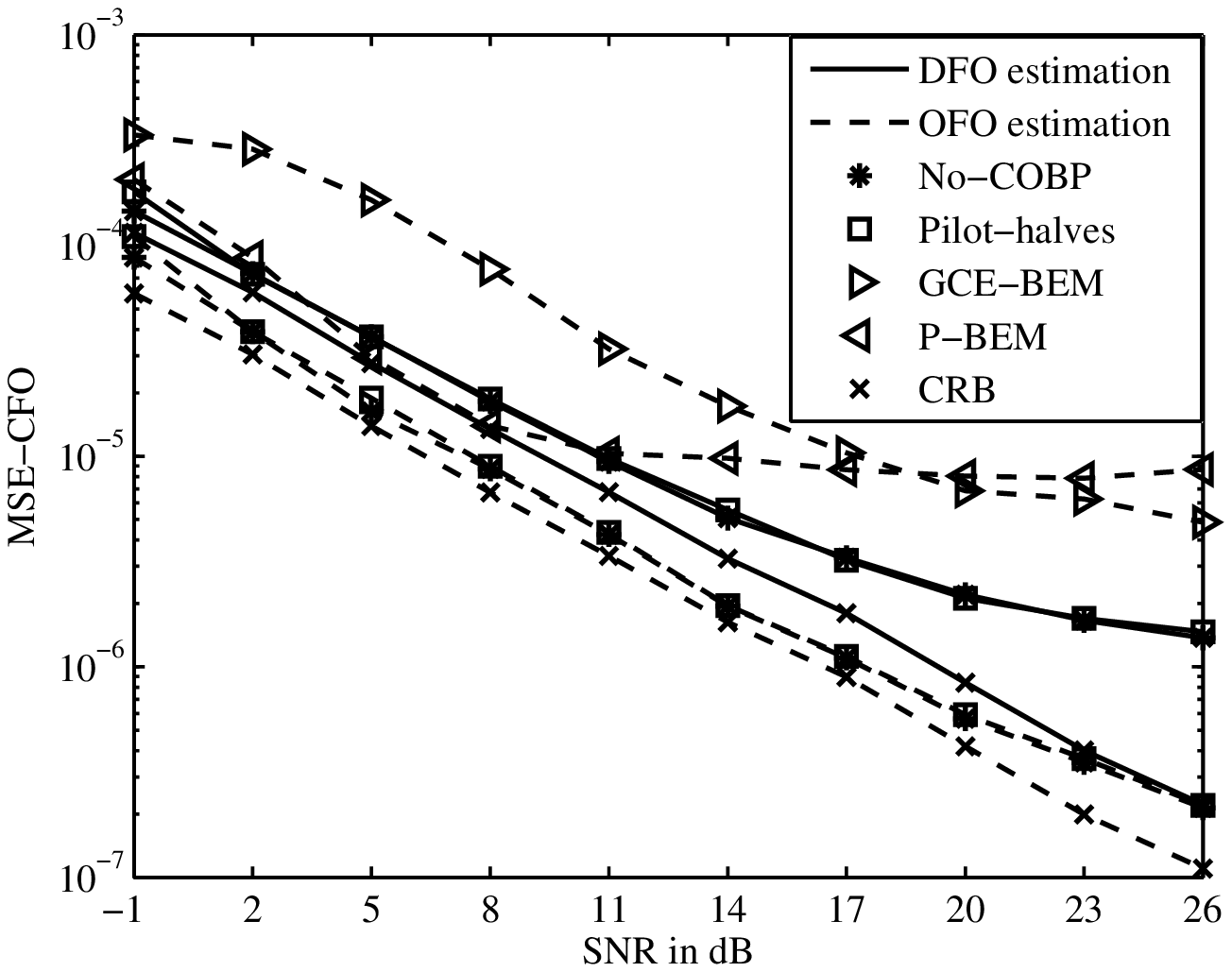}
    \caption{ MSE performance comparison of `No-COBP', `Pilot-halves', `GCE-BEM' and `P-BEM' with fully calibrated ULA ($K=1$) at $f_d=0.1$. }
  \end{minipage}
  \begin{minipage}{80mm}
  \centering
    \includegraphics[width=70mm]{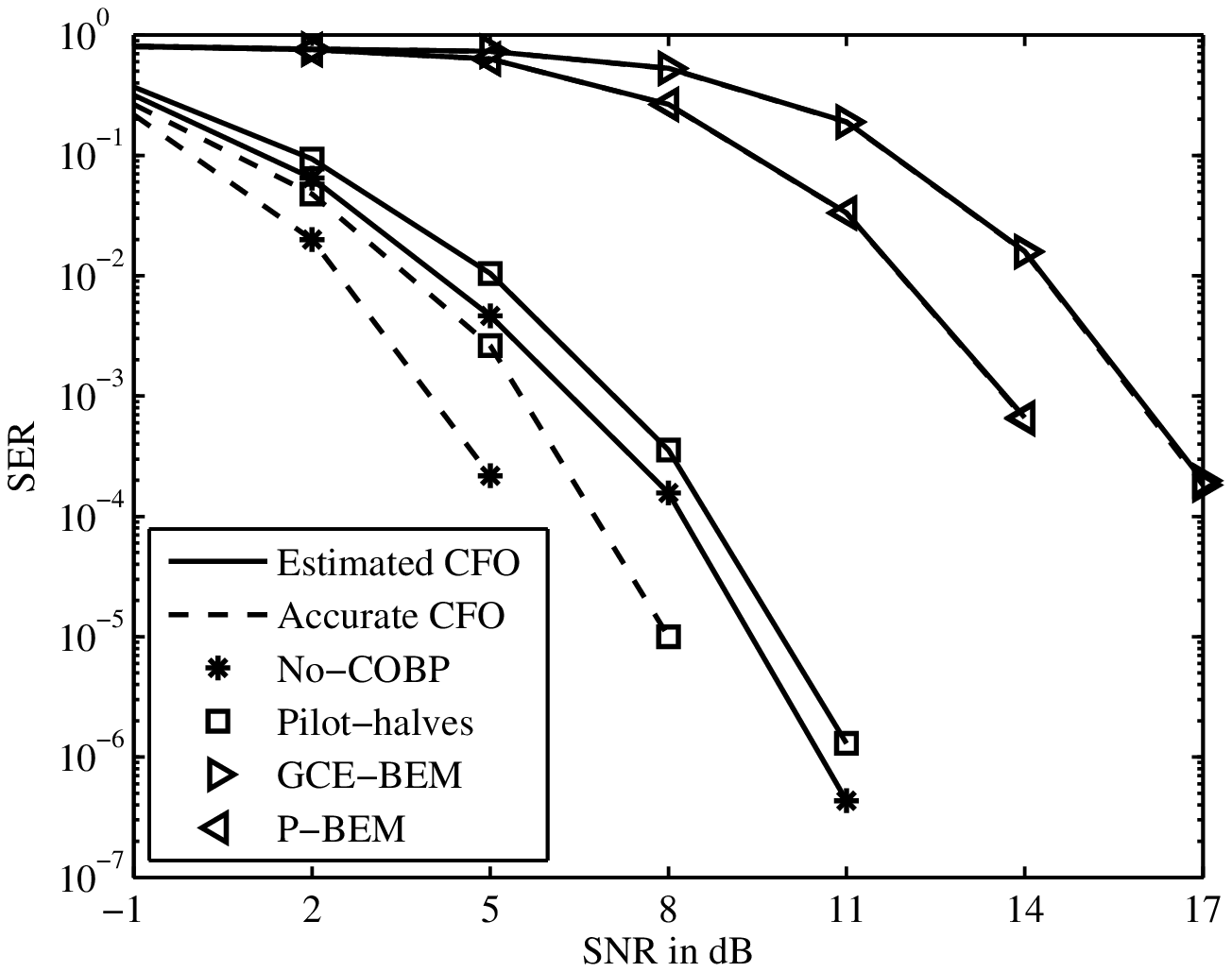}
    \caption{ SER performance comparison of `No-COBP', `Pilot-halves', `GCE-BEM' and `P-BEM' with fully calibrated ULA ($K=1$) at $f_d=0.1$. }
  \end{minipage}
\end{figure}

In Fig. 5 and Fig. 6, we assess the performance of `No-COBP' against the existing methods, including the scheme in~\cite{W_Guo2017TVT} (referred to as `Pilot-halves') and the most frequently encountered BEM approaches `GCE-BEM'~\cite{H_NguyenLe2010TB} and `P-BEM'~\cite{H_Hijazi2009TVT} (the first block and last block of each frame serve as pilot) under fully calibrated ULA ($K\!=\!1$). The nDFOmax remains $f_d \!=\! 0.1$.
Both figures corroborate the superiority of `No-COBP' over BEM. BEM exhibits obvious OFO estimation MSE floor, and compared to `No-COBP', the performance gaps of about 6dB and 8dB can be observed for `P-BEM' and `GCE-BEM'.
It is also observed that although `Pilot-halves' can achieve comparable MSE performance as `No-COBP', there is an SER performance gap of about 0.5dB. This can be attributed to the fact that the two-halves pilot exploited in~\cite{W_Guo2017TVT} exhibits some sparsity in the frequency domain, which may not be preferred for channel estimation and subsequent data detection~\cite{H_Minn2006TC}. However, the proposed algorithm enables the use of general pilot structure, which is more likely to provide superior detection performance.

\begin{figure}[t]
\setlength{\abovecaptionskip}{-0.5cm}
\setlength{\belowcaptionskip}{-1.05cm}
\begin{center}
\includegraphics[width=140mm]{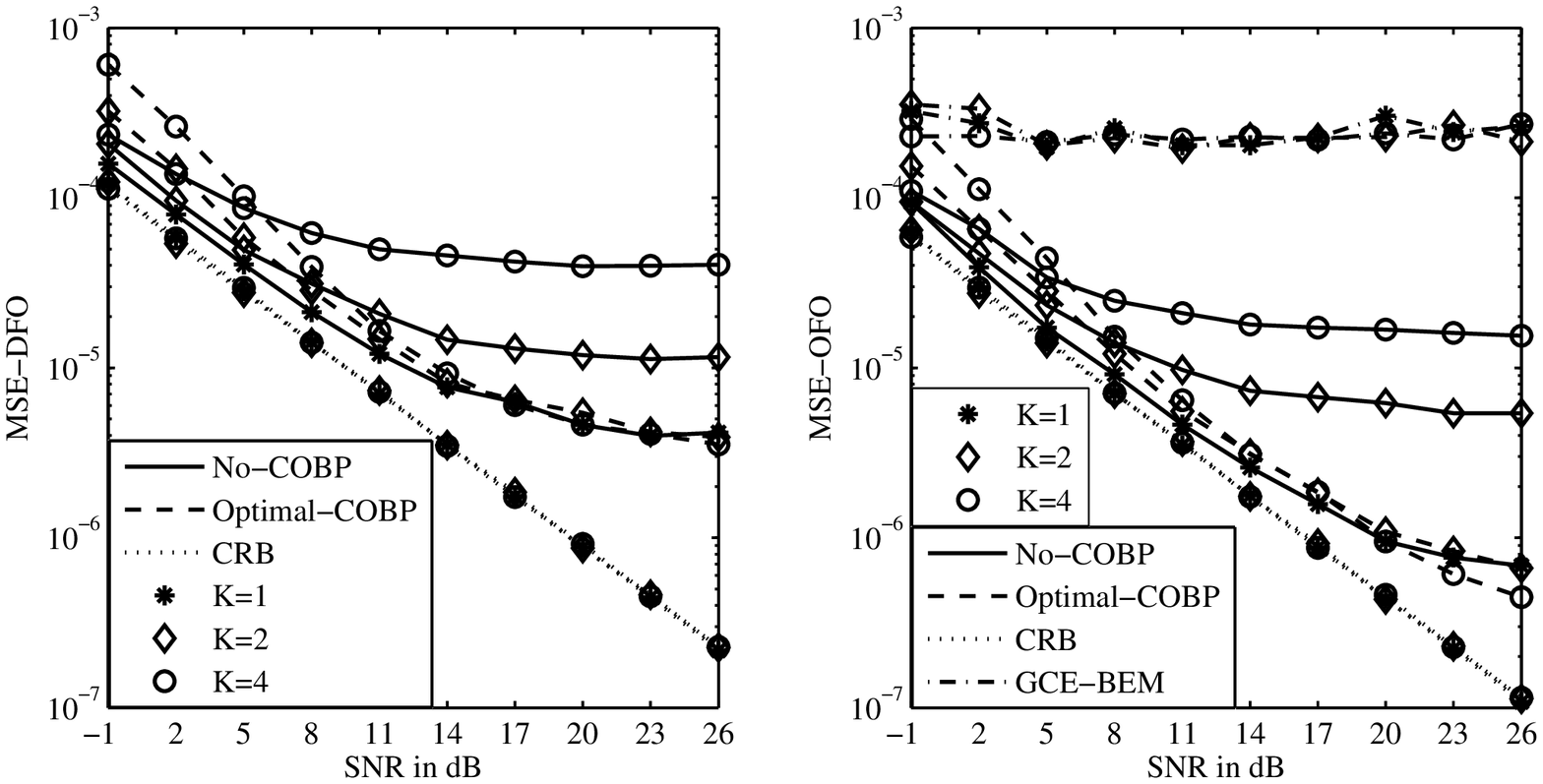}
\end{center}
\caption{ CFO estimation performance comparison of `No-COBP', `Optimal-COBP' and `GCE-BEM' with different numbers of subarrays ($f_d=0.4$, $K=1,2,4$). }
\end{figure}

\begin{figure}[t]
\setlength{\abovecaptionskip}{-0.5cm}
\setlength{\belowcaptionskip}{-1.2cm}
\begin{center}
\includegraphics[width=70mm]{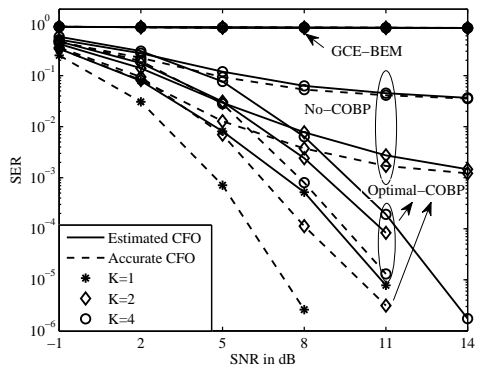}
\end{center}
\caption{ SER performance comparison of `No-COBP', `Optimal-COBP' and `GCE-BEM' with different numbers of subarrays ($f_d=0.4$, $K=1,2,4$). }
\end{figure}

In Fig. 7 and Fig. 8, we evaluate the CFO estimation and data detection performance of `No-COBP', `Optimal-COBP' and `GCE-BEM' with different numbers of subarrays $K=1, 2, 4$. Note that `No-COBP' and `Optimal-COBP' become identical at $K=1$.
From Fig. 7, we observe that:
1) Although insensitive to the inter-subarray mismatches, `GCE-BEM' suffers from high OFO estimation error floor.
2) The MSE performance of `No-COBP' degrades severely and drastically as the number of subarrays increases and the MSE floor is evident in the case of partly calibrated ULA. On the contrary, `Optimal-COBP' exhibits strong robustness to the number of subarrays.
3) The MSE performance of `Optimal-COBP' noticeably outperforms that of `No-COBP' at moderate and high SNRs, whereas the latter achieves better performance at low SNR. In fact, the system performance is mainly array mismatches-constrained at high SNR, and undoubtedly `Optimal-COBP' outperforms `No-COBP' since the former mitigates the impact of array mismatches with COBP. However, the system performance is noise-constrained at low SNR and thus `No-COBP' with fewer parameters to be estimated than `Optimal-COBP' will be superior.
4) The CRB obtained for different numbers of subarrays almost coincide, and the reason could be explained as follows. On the one hand, more estimation parameters would increase CRB. On the other hand, mismatches across more subarrays could enhance antenna diversities and thereby improve CRB. These two factors appear to offset each other approximately. In fact, the numerical MSEs of `Optimal-COBP' under different number of subarrays also asymptotically converge at high SNR. This also proves the effectiveness of COBP in mitigating the detrimental effects of inter-subarray gain and phase mismatches.

The results in Fig. 8 indicate that: 1) `GCE-BEM' fails to achieve reliable data detection for nDFOmax as large as $f_d=0.4$. 2) As expected, the SER performance of both `No-COBP' and `Optimal-COBP' relies on the number of subarrays, while the performance degradation of the former is much severer with the increase of number of subarrays. 3) There is an SER performance gap of about 2 or 3$\rm{dB}$ between `Optimal-COBP' and corresponding ideal case. 4) Although in contrast to the fully calibrated case, `Optimal-COBP' suffers from certain SER performance deterioration at a large number of subarrays (around 2dB at $K=4$), it still provides a feasible and so far best solution in the presence of inter-subarray mismatches. Even at $K=4$, the SER performance floor is not observed in this example.

\begin{figure}[htbp]
\vspace{-1.7em}
\setlength{\abovecaptionskip}{-0.1cm}
\setlength{\belowcaptionskip}{-0.9cm}
  \centering
  \begin{minipage}{80mm}
  \centering
    \includegraphics[width=70mm]{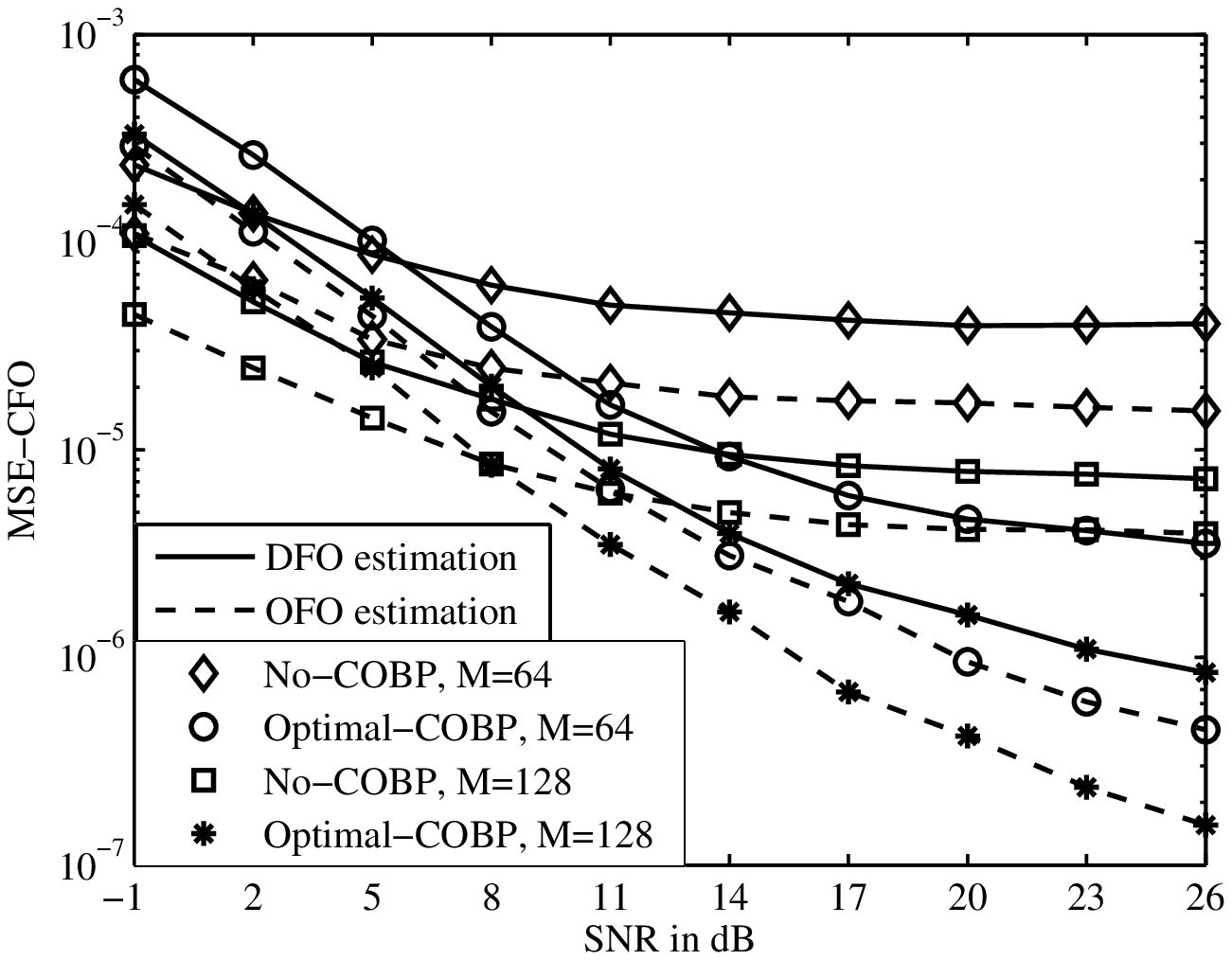}
    \caption{ MSE performance comparison of `No-COBP' and `Optimal-COBP' at different numbers of receive antennas ($f_d = 0.4, K=4, M = 64, 128$). }
  \end{minipage}
  \begin{minipage}{80mm}
  \centering
    \includegraphics[width=70mm]{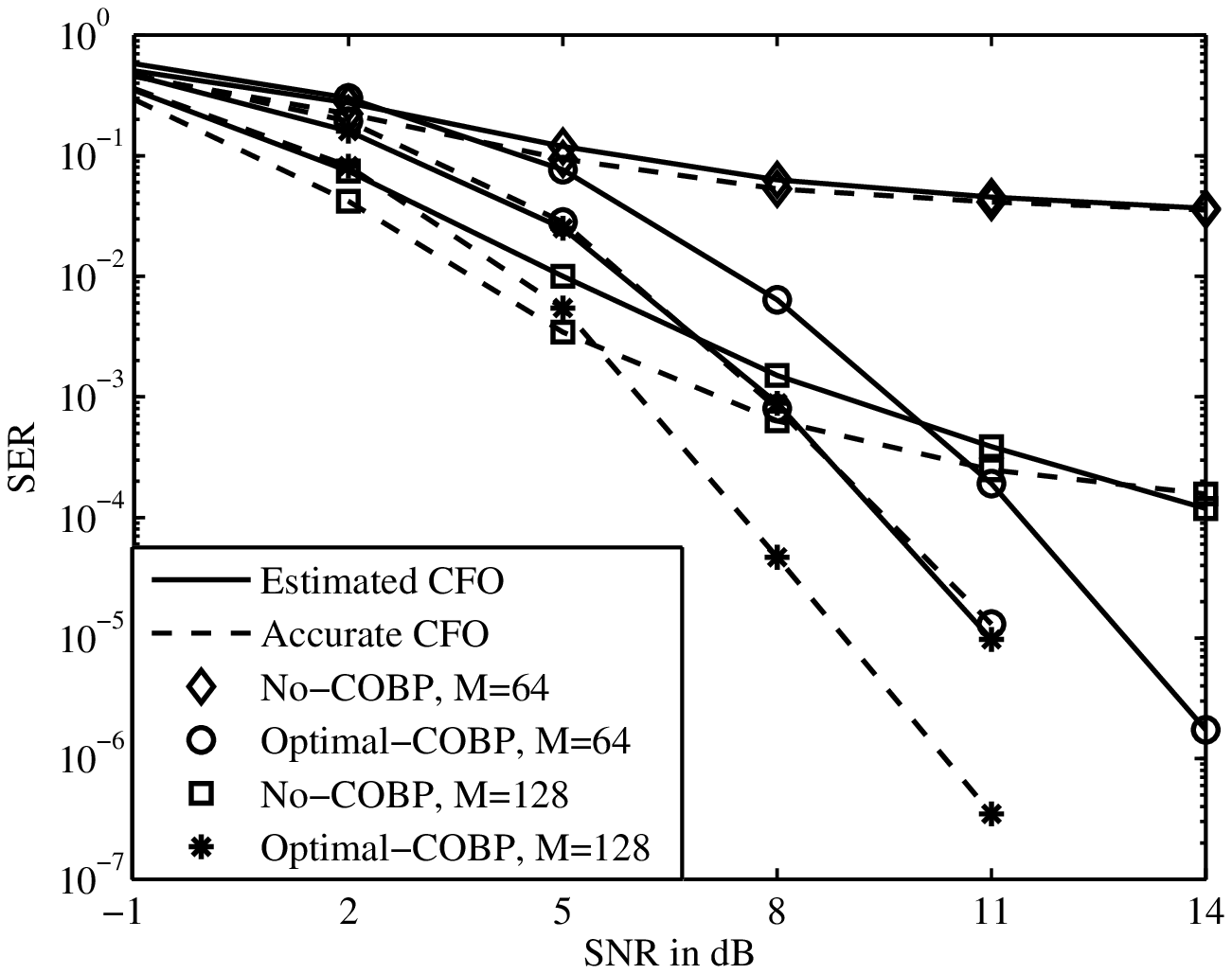}
    \caption{ SER performance comparison of `No-COBP' and `Optimal-COBP' at different numbers of receive antennas ($f_d = 0.4, K=4, M = 64, 128$). }
  \end{minipage}
\end{figure}

Next, the performance of `No-COBP' and `Optimal-COBP' are examined for $M\!=\!64$ and $128$ receive antennas in Fig. 9 and Fig. 10.
Though increasing $M$ from $64$ to $128$ effectively enhances the MSE and SER performances of both `No-COBP' and `Optimal-COBP', the former still suffers from visible CFO estimation error floor and high SER performance floor even under $M\!=\!128$. This further demonstrates the superiority of `Optimal-COBP' over `No-COBP' in the case of partly calibrated ULA.
Moreover, the following observations could be made: 1) In spite of the MSE performance floor, `No-COBP' indeed can provide valid coarse CFO estimates for `Optimal-COBP'.
2) If appropriately exploited, $128$ receive antennas should double the signal power at the receiver vis-a-vis $64$ antennas. Nonetheless, regarding the SER performance, it is observed that the array gain~\cite{A_Paulraj2003} (average increase of SNR at the receiver) is less than $3 \text{dB}$. In fact, even though the CFO estimation can be sufficiently accurate, the unwanted signals from undesired adjacent directions incorporated by a beamforming branch cannot be totally compensated, due to limited number of antennas. As a result of such incomplete compensation of frequency mismatch, the receiver is unable to achieve the thoroughly coherent combining of signals from different branches with different amplitudes and phases. Thus, the array gain which arises from the coherent combining effect of multiple antennas at the receiver (or transmitter or both) could not be fully exploited.

\begin{figure}[htbp]
\vspace{-1.5em}
\setlength{\abovecaptionskip}{-0.1cm}
\setlength{\belowcaptionskip}{0.5cm}
  \centering
  \begin{minipage}{80mm}
  \centering
    \includegraphics[width=70mm]{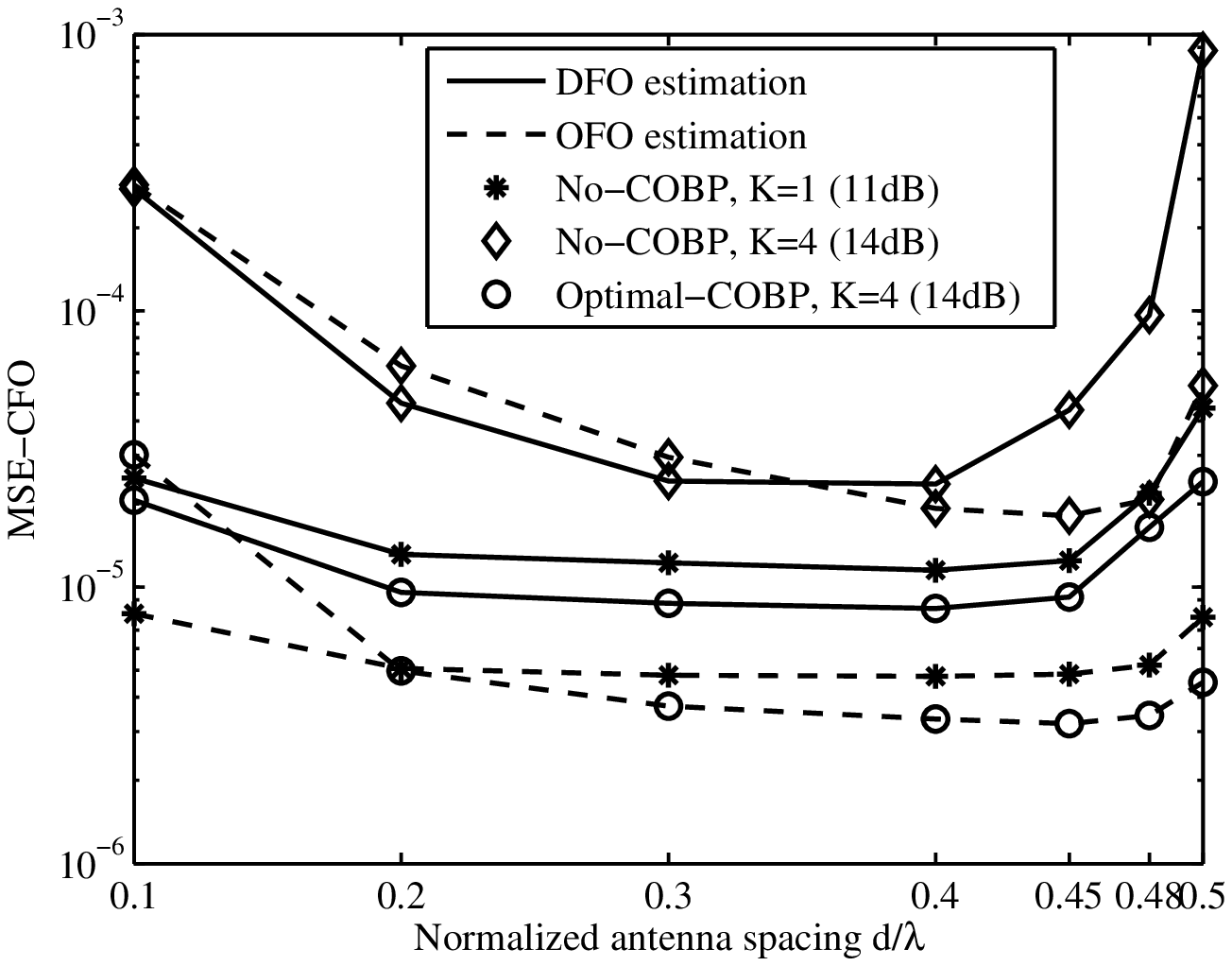}
    \caption{ MSE performance comparison of `No-COBP' and `Optimal-COBP' at different antenna spacings ($\tilde{d} = 0.1, 0.2, 0.3, 0.4, 0.45, 0.48, 0.5$). }
  \end{minipage}
  \begin{minipage}{80mm}
  \centering
    \includegraphics[width=70mm]{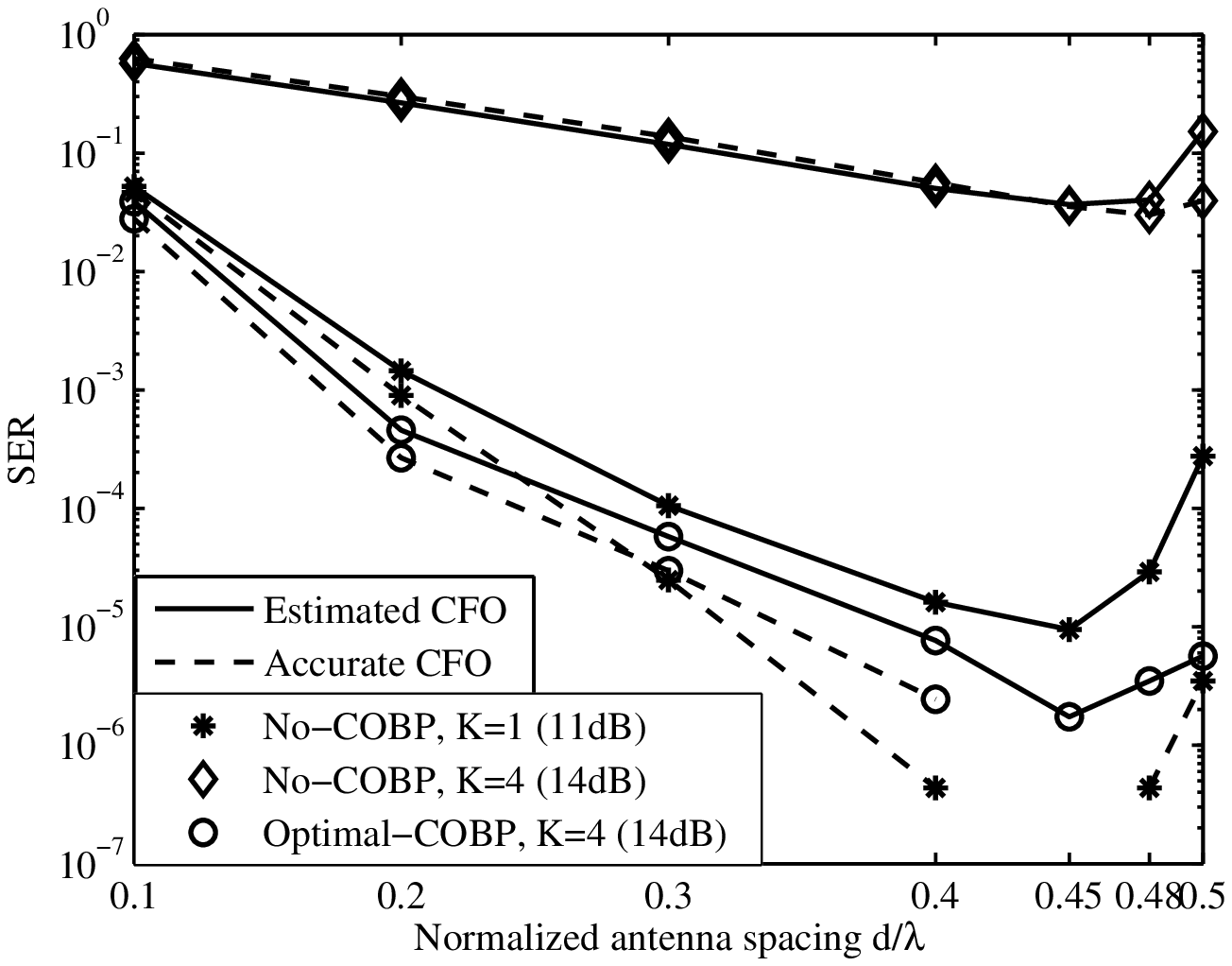}
    \caption{ SER performance comparison of `No-COBP' and `Optimal-COBP' at different antenna spacings ($\tilde{d} = 0.1, 0.2, 0.3, 0.4, 0.45, 0.48, 0.5$). }
  \end{minipage}
\end{figure}

At last, we gauge the performance of `No-COBP' and `Optimal-COBP' at different normalized antenna spacings $\tilde{d} = \{0.1, 0.2, 0.3, 0.4, 0.45, 0.48, 0.5\}$. The SNR is set as $14 \rm{dB}$ for $K=4$ and $11 \rm{dB}$ for $K=1$. Fig. 12 reveals explicitly the strong dependence of data detection performance on the antenna spacing, to which the CFO estimation performance is less sensitive for a wide range as indicated by Fig. 11. As predicated previously, on the one hand, too small antenna spacing (like $\tilde{d} = 0.1$ or $0.2$) cannot extract most of the spatial resolution provided by large-scale antenna array; on the other hand, an antenna spacing as large as $\tilde{d}=0.5$ ineluctably leads to aliasing between $\theta = 0^\circ$ and $\theta = 180^\circ$. Both cases are reflected by the SER performance exacerbation shown in Fig. 12. Such result justifies the empirical choice of $\tilde{d} = 0.45$.

\begin{table}[!t]
\caption{ Computational complexities of `No-COBP', `Optimal-COBP' and BEM }\label{CC_3Algorithms}
\vspace{1.8em}
\centering
\begin{tabular}{|c||c|c|}
\hline
Algorithms & \multicolumn{2}{|c|}{Computational complexities} \\
\hline
\multirow{4}*{`No-COBP'} & CFO estimation & $O\left( L\left( {{N\!+\!L}} \right)^{2} \!+\! \kappa QN\left( {{\log }_{2}}N\!+\!3N\!+\!9 \right) \right)$ \\
\cline{2-3}
$~ $ & \!Channel estimation\! & $O\left( QN\left( L\!+\!1 \right) \right)$ \\
\cline{2-3}
$~ $ & Data detection & $O\left( QN\left( {{N}_{b}}{{\log }_{2}}N\!+\!{{N}_{b}}\!+\!1 \right) \right)$ \\
\cline{2-3}
$~ $ & \multicolumn{2}{|c|}{$O\left( L{{\left( N\!+\!L \right)}^{2}}\!+\!\kappa QN\left( {{\log }_{2}}N\!+\!3N\!+\!9 \right)\!+\!QN\left( {{N}_{b}}{{\log }_{2}}N\!+\!{{N}_{b}}\!+\!L\!+\!2 \right) \right)$} \\
\hline
\multirow{5}*{\!`Optimal-COBP'\!} & CFO estimation & \!$O\left(
   L{{\left( N\!+\!L \right)}^{2}}\!+\!QN\left( K\left( M\!+\!4N\!+\!4K\!+\!5 \right)\!+\!M \right)
   \!+\!K\left( 2QK\!+\!2Q\!+\!{{K}^{2}} \right)
\right)$\! \\
\cline{2-3}
$~ $ & \!Channel estimation\! & $O\left( QN\left( K\!+\!L\!+\!1 \right) \right)$ \\
\cline{2-3}
$~ $ & Data detection & $O\left( QN\left( {{N}_{b}}{{\log }_{2}}N\!+\!\left( {{N}_{b}}\!-\!1 \right)K\!+\!{{N}_{b}}\!+\!1 \right) \right)$ \\
\cline{2-3}
$~ $ & \multicolumn{2}{|c|}{$O\left( \begin{matrix}
   L{{\left( N\!+\!L \right)}^{2}} \!+\!QNK\left( M\!+\!4N\!+\!4K\!+\!{{N}_{b}}\!+\!5 \right) \\
+QN\left( M\!+\!{{N}_{b}}{{\log }_{2}}N\!+\!L\!+\!{{N}_{b}}\!+\!2 \right) \!+\!K\left( 2QK\!+\!2Q\!+\!{{K}^{2}} \right)  \\
\end{matrix} \right)$} \\
\hline
\multirow{5}*{BEM} & OFO estimation & $O\left( LR{{\left( {{N}_{2}}\!+\!LR \right)}^{2}}\!+\!LR{{N}_{2}}\!+\!\kappa {{N}_{2}}\left( M\!+\!3{{N}_{2}}\!+\!8 \right) \right)$ \\
\cline{2-3}
$~ $ & \!Channel estimation\! & $O\left( MLR\left( {{N}_{2}}\!+\!N\left( {{N}_{b}}\!-\!1 \right) \right)\!+\! 2{{N}_{2}}M \right)$ \\
\cline{2-3}
$~ $ & Data detection & $O\left( \left( {{N}_{b}}\!-\!1 \right)\left( MN\left( 2{{N}^{2}}\!+\!N\!+\!1 \right)\!+\!{{N}^{3}} \right)\!+\!N{{\log }_{2}}N \right)$ \\
\cline{2-3}
$~ $ & \multicolumn{2}{|c|}{$O\left( \begin{matrix}
   LR{{\left( 2N\!+\!LR \right)}^{2}}\!+\!MN\left( LR\left( {{N}_{b}}\!+\!1 \right)\!+\!\left( 2{{N}^{2}}\!+\!N\!+\!1 \right)\left( {{N}_{b}}\!-\!1 \right)\!+\!4 \right)  \\
   \!+2\kappa N\left( M\!+\!6N\!+\!8 \right)\!+\!{{N}^{3}}\left( {{N}_{b}}\!-\!1 \right)\!+\!2NLR\!+\!N{{\log }_{2}}N  \\
\end{matrix} \right)$ } \\
\hline
\end{tabular}
\vspace{-2.5em}
\end{table}
Now, we will evaluate the computational complexities of not only CFO estimation, but also channel estimation and data detection of the proposed algorithms in comparison with BEM. The complexities in terms of complex multiplications of `No-COBP', `Optimal-COBP' and BEM are given in Table~\ref{CC_3Algorithms}. Here, $N_2 \!=\! 2N$, $\kappa$ denotes the number of iterations for CFO estimation, and $R$ is the order of basis functions used by BEM.
Consider $N\!=\!64, M\!=\!64, Q \!=\!64, L \!=\!8, N_b\!=\!4, K\!=\!4, \kappa\!=\!3, R \!=\!3$. The required complexities of the three algorithms are given in Table~\ref{CC_Comparison}. Apparently, the proposed algorithms `No-COBP' and `Optimal-COBP' profit from substantially reduced computational burdens than BEM approaches.

\section{Conclusion}
In this paper, we have addressed the joint estimation issue of DFO and OFO in high-mobility OFDM downlink in richly scattered wireless environments. The fully or partly calibrated massive ULA was adopted at the terminal HST to separate multiple CFOs via array beamforming, such that the doubly selective fading channel could be decomposed into a set of parallel frequency-selective fading channels in the angle domain, each of which is affected by a single dominant CFO and can be facilely managed. The use of iterative Newton's method has greatly reduced the computational burden of the joint estimation procedure. The necessity of introducing COBP was justified by the MSE performance analysis and it could effectively overcome the detrimental effects of array mismatches. Simulation results corroborated our proposed scheme.

\begin{table}[!t]
\vspace{2.2em}
\caption{ Comparison of the computational complexities between `No-COBP', `Optimal-COBP' and BEM }\label{CC_Comparison}
\vspace{1.8em}
\centering
\begin{tabular}{|c||c|c|c|c|}
\hline
\multirow{2}*{Algorithms} & \multicolumn{4}{|c|}{Computational complexity} \\
\cline{2-5}
$~ $ & CFO estimation & Channel estimation & Data detection & Overall \\
\hline
`No-COBP' & $2.59\times 10^6$ & $3.69\times 10^4$ & $1.19\times 10^5$ & $2.74\times 10^6$ \\
\hline
`Optimal-COBP' & $5.89\times 10^6$ & $5.32\times 10^4$ & $1.68\times 10^5$ & $6.11\times 10^6$ \\
\hline
BEM & $2.27\times 10^8$ & $5.08\times 10^5$ & $1.02\times 10^8$ & $3.29\times 10^8$ \\
\hline
\end{tabular}
\vspace{-2.5em}
\end{table}
%

\appendices
\section{Derivation of $a_{11}^{0}, a_{12}^{0}, a_{11}^{\mathrm{n}}, a_{12}^{\mathrm{n}}, a_{21}, a_{22}, a_{23}$}{\label{MSEDerivation}}

\emph{1) Calculation of $a_{11}^{0}$ and $a_{12}^{0}$.}

Denoting $ \mathbf{P}_{1} = \mathbf{DP}_{\mathbf{B}}^{\bot }+\mathbf{P}_{\mathbf{B}}^{\bot }{{\mathbf{D}}^{H}} $, we have
\begin{align}
\frac{\partial {{g}_{0}}}{\partial \tilde{\xi }} & =\int_{0}^{\rm{\pi}} {{\mathbf{a}}^{T}}\left( {\tilde{\theta }} \right)\mathbf{Y}_{0}^{H}\mathbf{E}\left( {\tilde{\varphi }} \right) \mathbf{P}_{1} {{\mathbf{E}}^{H}}\left( {\tilde{\varphi }} \right){{\mathbf{Y}}_{0}}{{\mathbf{a}}^{*}}\left( {\tilde{\theta }} \right) \sin \tilde{\theta }d\tilde{\theta }.
\end{align}

Define ${{\mathbf{E}}_{b}}=\mathbf{E}\left( {{f}_{d}}\left( \cos \tilde{\theta }-\cos {{\theta }_{p}} \right) \right)$ and $\overset{\scriptscriptstyle\frown}{\varphi }={{f}_{d}}\cos \tilde{\theta }+\xi $. Then, there is
\begin{align}\label{a120_intermed}
  & a_{12}^{0} = {E{\left. \left\{ \frac{\partial {{g}_{0}}}{\partial \tilde{\xi }} \right\} \right|}_{ \boldsymbol{\tilde{\phi}} = \boldsymbol{\phi} }} = \int_{0}^{\rm{\pi}}{{\mathbf{a}}^{T}}\left( {\tilde{\theta }} \right)\mathbf{Y}_{0}^{H}\mathbf{E}\left( {\overset{\scriptscriptstyle\frown}{\varphi }} \right) \mathbf{P}_{1} {{\mathbf{E}}^{H}}\left( {\overset{\scriptscriptstyle\frown}{\varphi }} \right){{\mathbf{Y}}_{0}}{{\mathbf{a}}^{*}}\left( {\tilde{\theta }} \right)\sin \tilde{\theta }d\tilde{\theta } \nonumber \\
\approx & \int_{0}^{\rm{\pi}}{{\mathbf{a}}^{T}}\left( {\tilde{\theta }} \right)\sum\limits_{p=1}^{P}{{\mathbf{V}}^{\text{*}}}\left( {{\theta }_{p}} \right){{\boldsymbol{\alpha }}^{\text{*}}}\mathbf{h}_{p}^{H}{{\mathbf{B}}^{H}}{{\mathbf{E}}_{b}} \mathbf{P}_{1} \mathbf{E}_{b}^{H}\mathbf{B}{{\mathbf{h}}_{p}}{{\boldsymbol{\alpha }}^{T}}{{\mathbf{V}}^{T}}\left( {{\theta }_{p}} \right){{\mathbf{a}}^{*}}\left( {\tilde{\theta }} \right)\sin \tilde{\theta }d\tilde{\theta } \nonumber \\
\approx & \frac{1}{P}\sum\limits_{p=1}^{P}\int_{0}^{\rm{\pi}}\operatorname{tr}\left[ {{\mathbf{B}}^{H}}{{\mathbf{E}}_{b}} \mathbf{P}_{1} \mathbf{E}_{b}^{H}\mathbf{B} \mathbf{\Lambda} \right] {{\mathbf{a}}^{T}}\left( {\tilde{\theta }} \right){{\mathbf{V}}^{\text{*}}}\left( {{\theta }_{p}} \right){{\boldsymbol{\alpha }}^{\text{*}}}{{\boldsymbol{\alpha }}^{T}}{{\mathbf{V}}^{T}}\left( {{\theta }_{p}} \right){{\mathbf{a}}^{*}}\left( {\tilde{\theta }} \right)\sin \tilde{\theta }d\tilde{\theta } \nonumber \\
\approx & \frac{1}{{\rm{\pi}}}\int_{0}^{\rm{\pi}}\int_{0}^{\rm{\pi}} \underbrace{\operatorname{tr}\left[ {{\mathbf{B}}^{H}}{{\mathbf{E}}_{b}} {\mathbf{P}_{1}} \mathbf{E}_{b}^{H}\mathbf{B}\mathbf{\Lambda} \right]}_{\eta } {{\boldsymbol{\alpha }}^{T}}{{\mathbf{V}}^{T}}\left( {{\theta }_{p}} \right){{\mathbf{V}}^{*}}\left( {\tilde{\theta }} \right)\mathbf{1} {{\mathbf{1}}^{T}}{{\mathbf{V}}^{T}}\left( {\tilde{\theta }} \right){{\mathbf{V}}^{\text{*}}}\left( {{\theta }_{p}} \right){{\boldsymbol{\alpha }}^{\text{*}}} \sin \tilde{\theta }d\tilde{\theta }d{{\theta }_{p}},
\end{align}
where $\mathbf{\Lambda} = \operatorname{diag}\left( \sigma_{1}^{2}, \sigma_{2}^{2}, \cdots\!, \sigma_{L}^{2} \right)$ and $\operatorname{tr}\left( \mathbf{\Lambda} \right) = \sum \nolimits_{l=1}^{L} {\sigma_{l}^{2}} = 1$. Denote $\tilde{x}=\cos \tilde{\theta }-\cos {{\theta }_{p}}$ for short. We have
\begin{align*}
{{\left[ {{\mathbf{V}}^{T}}\left( {{\theta }_{p}} \right){{\mathbf{V}}^{*}}\left( {\tilde{\theta }} \right) \right]}_{kl}}= \sum\limits_{m=J\left( k-1 \right)}^{Jk-1}{{{e}^{-\text{j}2\chi\left( \cos \tilde{\theta }-\cos {{\theta }_{p}} \right)m}}}{{\delta }_{kl}}
={{e}^{-\text{j}\chi \left( J-1 \right)\tilde{x}}} {{e}^{-\text{j}2\chi J \left( k-1 \right)\tilde{x}}} \frac{\sin \left( \chi J\tilde{x} \right)}{\sin \left( \chi\tilde{x} \right)}{{\delta }_{kl}},
\end{align*}
which leads to
\begin{align} \label{Ab}
{{\boldsymbol{\alpha }}^{T}}{{\mathbf{V}}^{T}}\left( {{\theta }_{p}} \right){{\mathbf{V}}^{*}}\left( {\tilde{\theta }} \right)\mathbf{1}{{\mathbf{1}}^{T}}{{\mathbf{V}}^{T}}\left( {\tilde{\theta }} \right){{\mathbf{V}}^{\text{*}}}\left( {{\theta }_{p}} \right){{\boldsymbol{\alpha }}^{\text{*}}}={{\boldsymbol{\alpha }}^{T}}{{\mathbf{A}}_{b}}{{\boldsymbol{\alpha }}^{\text{*}}},
\end{align}
where ${\mathbf{A}}_{b}$ is a $K \times K$ matrix whose $(p,q)$th element is given by $[{\mathbf{A}}_{b}]_{p,q} =\frac{{{\sin }^{2}}\left( \chi J\tilde{x} \right)}{{{\sin }^{2}}\left( \chi\tilde{x} \right)} {{e}^{\text{j}2\chi J\tilde{x} \left( q-p \right)}}$.

Besides, from ${{\mathbf{B}}^{H}}\mathbf{B} = N {{\left( {{\mathbf{F}}^{H}}\operatorname{diag}\left( \mathbf{x} \right){{\mathbf{F}}_{L}} \right)}^{H}}{{\mathbf{F}}^{H}}\operatorname{diag}\left( \mathbf{x} \right){{\mathbf{F}}_{L}}
 \approx N\mathbf{F}_{L}^{H}\operatorname{diag}\left( E\left\{ \left\| \mathbf{x} \right\|_{2}^{2} \right\} \right){{\mathbf{F}}_{L}} =N{{\mathbf{I}}_{L}}$, $\eta = 2\Re \left\{ \operatorname{tr}\left[ {{\mathbf{B}}^{H}}{{\mathbf{E}}_{b}} \mathbf{DP}_{\mathbf{B}}^{\bot } \mathbf{E}_{b}^{H}\mathbf{B}\mathbf{\Lambda} \right] \right\}$ can be simplified as
\begin{align} \label{eta}
 & \eta =2\operatorname{tr}\left[ \Re \left\{ \mathbf{E}_{b}^{H}\mathbf{B}\mathbf{\Lambda}{{\mathbf{B}}^{H}}{{\mathbf{E}}_{b}}\mathbf{D} - \frac{1}{N}{{\mathbf{B}}^{H}}{{\mathbf{E}}_{b}}\mathbf{DB}{{\mathbf{B}}^{H}}\mathbf{E}_{b}^{H}\mathbf{B}\mathbf{\Lambda}  \right\} \right] \nonumber \\
 \approx & -\frac{2}{N}\operatorname{tr}\left[ \Re \left\{ \operatorname{tr}\left( {{\mathbf{E}}_{b}}\mathbf{D} \right){{\mathbf{I}}_{L}} \operatorname{tr}\left( \mathbf{E}_{b}^{H} \right){{\mathbf{I}}_{L}} \mathbf{\Lambda} \right\} \right] = -\frac{2}{N} \operatorname{tr}\left( \mathbf{\Lambda} \right) \Re \left\{ \operatorname{tr}\left( \mathbf{E}_{b}^{H} \right)\operatorname{tr}\left( {{\mathbf{E}}_{b}}\mathbf{D} \right) \right\}, \nonumber \\
= & \frac{\text{2}{ \rm{\pi} }}{{{N}^{2}}} \left( \frac{{{\sin}^{\text{2}}}\left( { \rm{\pi} }{{f}_{d}}\tilde{x} \right)\cos \left( \frac{{ \rm{\pi} }}{N}{{f}_{d}}\tilde{x} \right)}{{{\sin }^{3}}\left( \frac{{ \rm{\pi} }}{N}{{f}_{d}}\tilde{x} \right)}-\frac{N\sin \left( { \rm{\pi} }{{f}_{d}}\tilde{x} \right)\cos \left( { \rm{\pi} }{{f}_{d}}\tilde{x} \right)}{{{\sin }^{2}}\left( \frac{{ \rm{\pi} }}{N}{{f}_{d}}\tilde{x} \right)} \right) \nonumber \\
\approx & 2{ \rm{\pi} }N\sin \left( { \rm{\pi} }{{f}_{d}}\tilde{x} \right) \frac{\sin \left( { \rm{\pi} }{{f}_{d}}\tilde{x} \right)-{ \rm{\pi} }{{f}_{d}}\tilde{x}\cos \left( { \rm{\pi} }{{f}_{d}}\tilde{x} \right)}{{{\left( { \rm{\pi} }{{f}_{d}}\tilde{x} \right)}^{3}}} \approx 2{ \rm{\pi} }N\sin \left( { \rm{\pi} }{{f}_{d}}\tilde{x} \right)\left( \frac{1}{3}-\frac{{{\left( { \rm{\pi} }{{f}_{d}}\tilde{x} \right)}^{2}}}{30} \right).
\end{align}

By combining (\ref{a120_intermed}), (\ref{Ab}) and (\ref{eta}), $a_{12}^{0} \!\approx\! \frac{1}{{ \rm{\pi} }}\int_{0}^{{ \rm{\pi} }}{\int_{0}^{ \rm{\pi} } {{\eta }{{\boldsymbol{\alpha }}^{T}}{{\mathbf{A}}_{b}}{{\boldsymbol{\alpha }}^{\text{*}}}\sin \tilde{\theta }d\tilde{\theta }}d{{\theta }_{p}}}$ can be simplified as (\ref{a120}).
Similarly, $a_{11}^{0} \!=\! {{\left. \left\{ \frac{\partial {{g}_{0}}}{\partial {{{\tilde{f}}}_{d}}}  \right\} \right|}_{\boldsymbol{\tilde{\phi}} = \boldsymbol{\phi} }} \!\approx\! \frac{1}{{ \rm{\pi} }}\int_{0}^{ \rm{\pi} }{\int_{0}^{ \rm{\pi} }{\cos \tilde{\theta } {\eta }{{\boldsymbol{\alpha }}^{T}}{{\mathbf{A}}_{b}}{{\boldsymbol{\alpha }}^{\text{*}}}\sin \tilde{\theta }d\tilde{\theta }}d{{\theta }_{p}}}$ can be simplified as (\ref{a110}).

\emph{2) Calculation of $a_{21}$, $a_{22}$ and $a_{23}$.}

From the zero-order Taylor series expansion, there is ${{\mathbf{E}}_{b}}\approx {{\mathbf{I}}_{N}}$. Besides, we have $\left\| \mathbf{P}_{\mathbf{B}}^{\bot }{{\mathbf{D}}^{H}}\mathbf{B\Lambda} \right\|_{2}^{2} \\
\approx \operatorname{tr}\left[ {{\mathbf{B}}^{H}}\mathbf{D}{{\mathbf{D}}^{H}}\mathbf{B\Lambda} \right] - \frac{1}{N} \operatorname{tr}\left[ {{\mathbf{B}}^{H}}{{\mathbf{D}}^{H}}\mathbf{B}{{\mathbf{B}}^{H}}\mathbf{DB\Lambda} \right]
\approx \operatorname{tr}\left( \mathbf{\Lambda} \right) \operatorname{tr}\left( \mathbf{D}{{\mathbf{D}}^{H}} \right)-\frac{\operatorname{tr}\left( \mathbf{\Lambda} \right)}{N}\operatorname{tr}\left( {{\mathbf{D}}^{H}} \right)\operatorname{tr}\left( \mathbf{D} \right)
 = \frac{{{ \rm{\pi} }^{2}}}{3}\frac{{{N}^{2}}-1}{N} \approx \frac{{{ \rm{\pi} }^{2}}}{3}N$. Denote ${\mathbf{P}}_{2} =  {{\mathbf{D}}^{2}}\mathbf{P}_{\mathbf{B}}^{\bot }+\mathbf{P}_{\mathbf{B}}^{\bot }{{\mathbf{D}}^{2H}}+2\mathbf{DP}_{\mathbf{B}}^{\bot }{{\mathbf{D}}^{H}} $.
Then, similar to (\ref{a120_intermed}), there is
\begin{align}\label{a23}
{{a}_{23}}= & {{\left. E\left\{ \frac{{{\partial }^{2}}{{g}_{0}}}{\partial {{{\tilde{\xi }}}^{2}}} \right\} \right|}_{\boldsymbol{\tilde{\phi}} = \boldsymbol{\phi} }} = \int_{0}^{ \rm{\pi} }{{{\mathbf{a}}^{T}}\left( {\tilde{\theta }} \right)\mathbf{Y}_{0}^{H}\mathbf{E}\left( {\overset{\scriptscriptstyle\frown}{\varphi }} \right){{\mathbf{P}}_{2}}{{\mathbf{E}}^{H}}\left( {\overset{\scriptscriptstyle\frown}{\varphi }} \right){{\mathbf{Y}}_{0}}{{\mathbf{a}}^{*}}\left( {\tilde{\theta }} \right)\sin \tilde{\theta }d\tilde{\theta }} \nonumber \\
\approx & \frac{1}{{ \rm{\pi} }}\int_{0}^{ \rm{\pi} }\int_{0}^{ \rm{\pi} }\operatorname{tr}\left[ {{\mathbf{B}}^{H}}{{\mathbf{E}}_{b}}{{\mathbf{P}}_{2}}\mathbf{E}_{b}^{H}\mathbf{B\Lambda} \right] {{{\boldsymbol{\alpha }}^{T}}{{\mathbf{A}}_{b}}{{\boldsymbol{\alpha }}^{\text{*}}}\sin \tilde{\theta }d\tilde{\theta }}d{{\theta }_{p}} \nonumber \\
\approx & \frac{2}{{ \rm{\pi} }}\left\| \mathbf{P}_{\mathbf{B}}^{\bot }{{\mathbf{D}}^{H}}\mathbf{B\Lambda} \right\|_{2}^{2}\int_{0}^{ \rm{\pi} }{\int_{0}^{ \rm{\pi} }{{{\boldsymbol{\alpha }}^{T}}{{\mathbf{A}}_{b}}{{\boldsymbol{\alpha }}^{\text{*}}}\sin \tilde{\theta }d\tilde{\theta }}d{{\theta }_{p}}} \approx \frac{2{\rm{\pi}}N}{3} \int_{0}^{ \rm{\pi} }{\int_{0}^{ \rm{\pi} }{{{\boldsymbol{\alpha }}^{T}}{{\mathbf{A}}_{b}}{{\boldsymbol{\alpha }}^{\text{*}}}\sin \tilde{\theta }d\tilde{\theta }}d{{\theta }_{p}}}.
\end{align}

In the same way, we obtain $a_{21}$ and $a_{22}$ given in (\ref{a21}) and (\ref{a22}), respectively.

\emph{3) Calculation of $a_{11}^{\mathrm{n}}$ and $a_{12}^{\mathrm{n}}$.}

Before the calculation, we first introduce the following Lemma.
\newtheorem{lemma}{Lemma}
\begin{lemma}
For steering vector ${\mathbf{a}}\left( {\theta} \right)$ whose ${r}$th element is defined as ${{e}^{\text{j}2\chi\left( r-1 \right)\cos \theta }}$, there holds
\begin{align}
\int_{0}^{ \rm{\pi} }\int_{0}^{ \rm{\pi} } & { {{\mathbf{a}}^{T}}\left( {\bar{\theta }} \right) {{\mathbf{a}}^{*}}\left( {\tilde{\theta }} \right) f\left( {\bar{\theta }} \right)d\cos \tilde{\theta }}d\cos \bar{\theta }
 \approx \int_{0}^{ \rm{\pi} }{f\left( {\tilde{\theta }} \right) \frac{\sin \tilde{\theta }}{\tilde{d}} d\tilde{\theta }}.
\end{align}
\end{lemma}

\begin{IEEEproof}
Denote $u\left( \cdot  \right)$ as the unit step function and $g\left( x, x_0 \right)=\frac{\sin \left( \chi M\left( x-{{x}_{0}} \right) \right)}{\chi \left( x-{{x}_{0}} \right)}$. Then, its Fourier Transform is given by $G\left( \varpi  \right) =\mathscr{F}\{\frac{\sin \left( \chi M\left( x-{{x}_{0}} \right) \right)}{\chi \left( x-{{x}_{0}} \right)} \}=\frac{1}{\tilde{d}}{{e}^{-\text{j}\varpi {{x}_{0}}}}\left[ u\left( \varpi \!+\!\chi M \right)-u\left( \varpi \!-\!\chi M \right) \right]$.

Moreover, define $F\left( \varpi  \right)$ as the Fourier Transform of $f\left( x \right)$. Then, there holds
\begin{align}
  & \underset{M\to \infty }{\mathop{\lim }}\,\int_{-\infty }^{+\infty }{ g\left( x, x_0 \right) f\left( x \right)dx} = {{\left. \underset{M\to \infty }{\mathop{\lim }}\,\frac{1}{2{ \rm{\pi} }}G\left( \varpi  \right)\otimes F\left( \varpi  \right) \right|}_{\varpi =0}} \nonumber \\
= & \underset{M\to \infty }{\mathop{\lim }}\,\frac{1}{2{ \rm{\pi} }}\int_{-\chi M}^{\chi M}{\frac{1}{\tilde{d}}{{e}^{\text{j}\Omega {{x}_{0}}}}F\left( \Omega  \right)d\Omega } = \frac{1}{\tilde{d} 2{ \rm{\pi} }}\int_{-\infty }^{+\infty }{F\left( \Omega  \right){{e}^{\text{j}\Omega {{x}_{0}}}}d\Omega }=\frac{1}{\tilde{d}}f\left( {{x}_{0}} \right).
\end{align}

Therefore, we have
\begin{align}
  & \int_{0}^{ \rm{\pi} }{\int_{0}^{ \rm{\pi} }{ {{\mathbf{a}}^{T}}\left( {\bar{\theta }} \right) {{\mathbf{a}}^{*}}\left( {\tilde{\theta }} \right) f\left( {\bar{\theta }} \right)d\cos \tilde{\theta }}d\cos \bar{\theta }} \nonumber \\
= & \int_{1}^{-1}{\int_{1}^{-1}{\frac{\sin \left( \chi M\left( y-x \right) \right)}{\sin \left( \chi \left( y-x \right) \right)}{{e}^{\text{j}\chi \left( M-1 \right)\left( y-x \right)}}f\left( y \right)}dxdy} \left( x=\cos \tilde{\theta }, y=\cos \bar{\theta } \right) \nonumber \\
\approx & \int_{-1}^{1}{\int_{-\infty }^{\infty }{\frac{\sin \left( \chi M\left( y-x \right) \right)}{\chi \left( y-x \right)}{{e}^{\text{j}\chi \left( M-1 \right)\left( y-x \right)}}f\left( y \right)dy}dx} \nonumber \\
\approx & \frac{1}{\tilde{d}}\int_{-1}^{1}{f\left( x \right)dx} \ \overset{x=\cos \tilde{\theta }}{\mathop{=}}\, \ \frac{1}{\tilde{d}}\int_{0}^{ \rm{\pi} }{f\left( {\tilde{\theta }} \right)\sin \tilde{\theta }d\tilde{\theta }}.
\end{align}

This completes the proof.
\end{IEEEproof}

Denote ${\mathbf{P}}_{1}\left( {\tilde{\varphi }} \right) =  \mathbf{E}\left( {\tilde{\varphi }} \right)\left( \mathbf{DP}_{\mathbf{B}}^{\bot }+\mathbf{P}_{\mathbf{B}}^{\bot }{{\mathbf{D}}^{H}} \right){{\mathbf{E}}^{H}}\left( {\tilde{\varphi }} \right) $. Then
\begin{align*}
\frac{\partial {{g}_{\mathrm{n}}}}{\partial \tilde{\xi }} = -2\Re \left\{ \int_{0}^{ \rm{\pi} } {{{\mathbf{a}}^{T}}\left( {\tilde{\theta }} \right)\mathbf{Y}_{0}^{H}{\mathbf{P}}_{1}\left( {\tilde{\varphi }} \right)\mathbf{W}{{\mathbf{a}}^{*}}\left( {\tilde{\theta }} \right)d\cos \tilde{\theta }} \right\}.
\end{align*}

By virtue of Lemma 1, we arrive at
\begin{align}\label{Lemma1_usage}
  & E\left\{ {{\left( \frac{\partial {g_{\mathrm{n}}}}{\partial \tilde{\xi }} \right)}^{2}} \right\} = 2E\left\{ {{\left| \int_{0}^{ \rm{\pi} }{{{\mathbf{a}}^{T}}\left( {\tilde{\theta }} \right)\mathbf{Y}_{0}^{H}{\mathbf{P}}_{1}\left( {\tilde{\varphi }} \right) \mathbf{W}{{\mathbf{a}}^{*}}\left( {\tilde{\theta }} \right)d\cos \tilde{\theta }} \right|}^{2}} \right\} \nonumber \\
\approx & 2\sigma _{\mathrm{n}}^{2}\int_{0}^{ \rm{\pi} }\int_{0}^{ \rm{\pi} }
\operatorname{tr} \left[ {{\mathbf{a}}^{*}}\left( {\tilde{\theta }} \right) {{\mathbf{a}}^{T}}\left( {\bar{\theta }} \right) \right] \underbrace{{{\mathbf{a}}^{T}}\left( {\tilde{\theta }} \right)\mathbf{Y}_{0}^{H}{\mathbf{P}}_{1}\left( {\tilde{\varphi }} \right) {\mathbf{P}}_{1}\left( {\bar{\varphi }} \right) {{\mathbf{Y}}_{0}}{{\mathbf{a}}^{*}}\left( {\bar{\theta }} \right)}_{f\left( \bar{\theta} \right)} d\cos \bar{\theta }d\cos \tilde{\theta } \nonumber \\
 \approx & \frac{2\sigma _{\mathrm{n}}^{2}}{\tilde{d}}\int_{0}^{ \rm{\pi} }{{\mathbf{a}}^{T}}\left( {\tilde{\theta }} \right)\mathbf{Y}_{0}^{H}\mathbf{E}\left( {\tilde{\varphi }} \right)\mathbf{P} {{\mathbf{E}}^{H}}\left( {\tilde{\varphi }} \right){{\mathbf{Y}}_{0}}{{\mathbf{a}}^{*}}\left( {\tilde{\theta }} \right)\sin \tilde{\theta }d\tilde{\theta },
\end{align}
where $\mathbf{P} =\mathbf{DP}_{\mathbf{B}}^{\bot }{{\mathbf{D}}^{H}}+\mathbf{DP}_{\mathbf{B}}^{\bot }\mathbf{DP}_{\mathbf{B}}^{\bot }+\mathbf{P}_{\mathbf{B}}^{\bot }{{\mathbf{D}}^{H}}\mathbf{P}_{\mathbf{B}}^{\bot }{{\mathbf{D}}^{H}}+\mathbf{P}_{\mathbf{B}}^{\bot }{{\mathbf{D}}^{H}}\mathbf{DP}_{\mathbf{B}}^{\bot }$.

In the same way as (\ref{a120_intermed}) and (\ref{a23}), we obtain
\begin{align}
 & a_{12}^{\mathrm{n}} = E{{\left. \left\{ {{\left( \frac{\partial {g_{\mathrm{n}}}}{\partial \tilde{\xi }} \right)}^{2}} \right\} \right|}_{\boldsymbol{\tilde{\phi}} = \boldsymbol{\phi} }}
\approx \frac{2\sigma _{\mathrm{n}}^{2}}{\tilde{d}}\int_{0}^{ \rm{\pi} } {{\mathbf{a}}^{T}}\left( {\tilde{\theta }} \right)\mathbf{Y}_{0}^{H}\mathbf{E}\left( {\overset{\scriptscriptstyle\frown}{\varphi }} \right)\mathbf{P} {{\mathbf{E}}^{H}}\left( {\overset{\scriptscriptstyle\frown}{\varphi }} \right){{\mathbf{Y}}_{0}}{{\mathbf{a}}^{*}}\left( {\tilde{\theta }} \right)\sin \tilde{\theta }d\tilde{\theta } \nonumber \\
\approx & \frac{2\sigma _{\mathrm{n}}^{2}}{\tilde{d}{\rm{\pi}} }\left\| \mathbf{P}_{\mathbf{B}}^{\bot }{{\mathbf{D}}^{H}}\mathbf{B\Lambda} \right\|_{2}^{2} \int_{0}^{ \rm{\pi} }{\int_{0}^{ \rm{\pi} } {{{\boldsymbol{\alpha }}^{T}}{{\mathbf{A}}_{b}}{{\boldsymbol{\alpha }}^{*}}\sin \tilde{\theta } d\tilde{\theta }}d{{\theta }_{p}}} \approx \frac{2{\rm{\pi}}N \sigma_{\mathrm{n}}^{2}}{3\tilde{d}} \int_{0}^{ \rm{\pi} }{\int_{0}^{ \rm{\pi} } {{{\boldsymbol{\alpha }}^{T}}{{\mathbf{A}}_{b}}{{\boldsymbol{\alpha }}^{*}}\sin \tilde{\theta } d\tilde{\theta }}d{{\theta }_{p}}}.
\end{align}
Similarly, we can obtain $a_{11}^{\mathrm{n}}$ given by (\ref{a11n}).

\vspace{-0.6em}
\section{Demonstration of the negligibility of term $a_{22}$}{\label{CrossTerm}}
Let ${f_k}\left( \tilde{\theta },{{\theta }_{p}} \right)$ denote the following function
\begin{align}
& {f_k}\left( \tilde{\theta },{{\theta }_{p}} \right) = \sin \tilde{\theta }\frac{{{\sin }^{2}}\left( \chi J\left( \cos \tilde{\theta }-\cos {{\theta }_{p}} \right) \right)}{{{\sin }^{2}}\left( \chi \left( \cos \tilde{\theta }-\cos {{\theta }_{p}} \right) \right)}{{e}^{\text{j}2\chi J\left( \cos \tilde{\theta }-\cos {{\theta }_{p}} \right)k}}.
\end{align}

Define $\zeta _{22}^{k}=\int_{0}^{ \rm{\pi} }{\int_{0}^{ \rm{\pi} }{2\cos \tilde{\theta }{f_k}\left( \tilde{\theta },{{\theta }_{p}} \right)d\tilde{\theta }}d{{\theta }_{p}}}$. Then, it is not difficult to verify that
\begin{align}
  & \Re \left\{ \cos \left( { \rm{\pi} }-\tilde{\theta } \right){f_k}\left( { \rm{\pi} }-\tilde{\theta },{ \rm{\pi} }-{{\theta }_{p}} \right) \right\}=-\Re \left\{ \cos \tilde{\theta }{f_k}\left( \tilde{\theta },{{\theta }_{p}} \right) \right\}, \nonumber \\
 & \Re \left\{ \cos \left( { \rm{\pi} }-\tilde{\theta } \right){f_k}\left( { \rm{\pi} }-\tilde{\theta },{{\theta }_{p}} \right) \right\}=-\Re \left\{ \cos \tilde{\theta }{f_k}\left( \tilde{\theta },{ \rm{\pi} } -{{\theta }_{p}} \right) \right\}.
\end{align}

Therefore, we have $\Re \left\{ \zeta _{22}^{k} \right\}=\Re \left\{ \int_{0}^{ \rm{\pi} }{\int_{0}^{ \rm{\pi} }{2\cos \tilde{\theta }{f_k}\left( \tilde{\theta },{{\theta }_{p}} \right)d\tilde{\theta }}d{{\theta }_{p}}} \right\}=0$, $\zeta _{22}^{0}=\Re \left\{ \zeta _{22}^{0} \right\}=0$ and ${{\left. \zeta _{22}^{k} \right|}_{k\ne 0}}=\text{j}\Im \left\{ \zeta _{22}^{k} \right\}$.
For a given ${{\theta }_{p}}$, the range of $\tilde{\theta }$ constraining ${f_k}\left( \tilde{\theta },{{\theta }_{p}} \right)$ in the main beam lobe is determined by $\left| \cos \tilde{\theta }-\cos {{\theta }_{p}} \right|\le \frac{1}{\tilde{d} J}$, i.e.,
$\vartheta_{1} = \arccos \left( \cos {{\theta }_{p}}+\frac{1}{\tilde{d} J} \right)\le \tilde{\theta }\le \arccos \left( \cos {{\theta }_{p}}-\frac{1}{\tilde{d} J} \right) = \vartheta_{2} $.
Besides, $\frac{{{\sin }^{2}}{ \rm{\pi} }X\varphi }{{{\sin }^{2}}{ \rm{\pi} }\varphi }\lessapprox {{X}^{2}}{{\cos }^{2}}\frac{{ \rm{\pi} }X\varphi }{2}$ holds for $\left| \varphi  \right|\le \frac{1}{X}$.

Hence, there will be
\begin{align}\label{zeta22}
  & {{\left. \Im \left\{ \zeta _{22}^{k} \right\} \right|}_{k\ne 0}} =\Im \left\{ \int_{0}^{ \rm{\pi} }{\int_{0}^{ \rm{\pi} }{2\cos \tilde{\theta }{f_k}\left( \tilde{\theta },{{\theta }_{p}} \right)d\tilde{\theta }}d{{\theta }_{p}}} \right\}, \nonumber \\
\approx & -{J^{2}}\int_{0}^{ \rm{\pi} }\int_{\vartheta_{1}}^{\vartheta_{2}}\text{2}\cos \tilde{\theta } {{\cos }^{2}}\left( \frac{\chi J}{2}\left( \cos \tilde{\theta }-\cos {{\theta }_{p}} \right) \right) \sin \left( 2\chi J\left( \cos \tilde{\theta }-\cos {{\theta }_{p}} \right)k \right)d\cos \tilde{\theta }d{{\theta }_{p}} \nonumber \\
 \overset{*}{\mathop{=}}\, & \frac{2}{{{\chi}^{2}}}\int_{0}^{ \rm{\pi} } {\int_{-{ \rm{\pi} }}^{ \rm{\pi} }{\left( \tilde{y}+\chi J\cos {{\theta }_{p}} \right){{\cos }^{2}}\frac{{\tilde{y}}}{2}\sin \left( 2k\tilde{y} \right)d\tilde{y}}d{{\theta }_{p}}} \nonumber \\
= & \frac{\pi}{2{{\chi}^{2}}} \int_{-{ \rm{\pi} }}^{ \rm{\pi} }\tilde{y} \left( 2\sin \left( 2k\tilde{y} \right)+\sin \left( \left( 2k+1 \right)\tilde{y} \right)+\sin \left( \left( 2k-1 \right)\tilde{y} \right) \right)d\tilde{y} = \frac{1}{{{\tilde{d}}^{2}}k\left( 4{{k}^{2}}-1 \right)},
\end{align}
where $\overset{*}{\mathop{=}}$ in (\ref{zeta22}) and hereinbelow is the marker of variable substitution ${\tilde{y}=\chi J\left( \cos \tilde{\theta }-\cos {{\theta }_{p}} \right)}$.

Similarly, define $\zeta _{21}^{k}=\int_{0}^{ \rm{\pi} }{\int_{0}^{ \rm{\pi} }{{{\cos }^{2}}\tilde{\theta }{{f}_{k}}\left( \tilde{\theta },{{\theta }_{p}} \right)d\tilde{\theta }}d{{\theta }_{p}}}$. Then, we have $\zeta _{21}^{k}=\Re \left\{ \zeta _{21}^{k} \right\}$ and
\begin{align}\label{zeta21}
 \zeta _{21}^{0}= & \int_{0}^{ \rm{\pi} }{\int_{0}^{ \rm{\pi} }{{{\cos }^{2}}\tilde{\theta }{{f}_{0}}\left( \tilde{\theta },{{\theta }_{p}} \right)d\tilde{\theta }}d{{\theta }_{p}}}
 \approx -{{J}^{2}}\int_{0}^{ \rm{\pi} }{\int_{\vartheta_{1}}^{\vartheta_{2}}{{{\cos }^{2}}\tilde{\theta }}} {{\cos }^{2}}\left( \frac{\chi J}{2}\left( \cos \tilde{\theta }-\cos {{\theta }_{p}} \right) \right)d\cos \tilde{\theta }d{{\theta }_{p}} \nonumber \\
 \overset{*}{\mathop{=}}\,& \frac{J}{\chi}\int_{0}^{ \rm{\pi} }{\int_{-{ \rm{\pi} }}^{ \rm{\pi} }{{{\left( \frac{{\tilde{y}}}{\chi J}+\cos {{\theta }_{p}} \right)}^{2}}\frac{\cos \tilde{y}+1}{2}d\tilde{y}}d{{\theta }_{p}}}
\approx  \frac{J}{\text{2}\chi}\int_{0}^{ \rm{\pi} }{{{\cos }^{2}}{{\theta }_{p}}\int_{-{ \rm{\pi} }}^{ \rm{\pi} }{\left( \cos \tilde{y}+1 \right)d\tilde{y}}d{{\theta }_{p}}}=\frac{{ \rm{\pi} }J}{\text{2}\tilde{d}}.
\end{align}

Besides, define $\zeta _{23}^{k}=\int_{0}^{ \rm{\pi} }{\int_{0}^{ \rm{\pi} }{{{f}_{k}}\left( \tilde{\theta },{{\theta }_{p}} \right)d\tilde{\theta }}d{{\theta }_{p}}}$. Then, we have $\zeta _{23}^{k}=\Re \left\{ \zeta _{23}^{k} \right\}$ and similar to the derivation of (\ref{zeta21}), there holds
\begin{align}\label{zeta23}
  \zeta _{23}^{0}=\int_{0}^{ \rm{\pi} }{\int_{0}^{ \rm{\pi} }{{{f}_{0}}\left( \tilde{\theta },{{\theta }_{p}} \right)d\tilde{\theta }}d{{\theta }_{p}}} \approx \frac{{ \rm{\pi} }J}{\tilde{d}}.
\end{align}

The results in (\ref{zeta22}), (\ref{zeta21}) and (\ref{zeta23}) reveal that $\zeta _{23}^{0}\approx 2\zeta _{21}^{0}\gg \left| \zeta _{22}^{k} \right|$. Taking $\tilde{d}=0.45,\ M=64$ and $K=4$ for example, we  have $\zeta _{21}^{0}\approx 55.85,\ \zeta _{23}^{0}\approx 111.7,\ \zeta _{22}^{0}=0,\ \zeta _{22}^{1} = {\zeta _{22}^{-1}}^{*}\approx 1.65\text{j},\ \zeta _{22}^{2} = {\zeta _{22}^{-2}}^{*} \approx 0.165\text{j},\ \zeta _{22}^{3} = {\zeta _{22}^{-3}}^{*} \approx 0.047\text{j}$.

Moreover, define ${\mathbf{A}}_{2n}$ whose $(p,q)$th element is $[{\mathbf{A}}_{2n}]_{p,q} = \zeta _{2n}^{q-p}$. We have ${{a}_{2n}}=\frac{2{\rm{\pi}}N}{3}{{\boldsymbol{\alpha }}^{T}} \mathbf{A}_{2n} {{\boldsymbol{\alpha }}^{\text{*}}}, \\ n=1,2,3 $.
Thus, there holds ${{a}_{23}} \approx 2{{a}_{21}}\gg {{a}_{22}}$ and ${a}_{22}$ is negligible compared to ${{a}_{21}}$ and ${{a}_{23}}$.

\vspace{-0.6em}
\section{MSE performance simplification in the case of fully calibrated ULA}{\label{MSE_ULA_proof}}
From (\ref{MSE0n}), there holds
\begin{align}\label{MSEn_pULA}
{{\text{MSE}}_{\mathrm{n}}}\left\{ {{{\tilde{f}}}_{d}} \right\} & \approx \frac{3\sigma _{\mathrm{n}}^{2}}{2{ \rm{\pi} }N\tilde{d} \underbrace{\int_{0}^{ \rm{\pi} }{\int_{0}^{ \rm{\pi} }{{{\cos }^{2}}\tilde{\theta }{{\boldsymbol{\alpha }}^{T}}{{\mathbf{A}}_{b}}{{\boldsymbol{\alpha }}^{\text{*}}} \sin \tilde{\theta} d\tilde{\theta }}d{{\theta }_{p}}}}_{{\rho_{1}}} }, \nonumber \\
{{\text{MSE}}_{\mathrm{n}}}\left\{ {{{\tilde{\xi}}}} \right\} & \approx \frac{3\sigma _{\mathrm{n}}^{2}}{2{ \rm{\pi} }N\tilde{d} \underbrace{\int_{0}^{ \rm{\pi} }{\int_{0}^{ \rm{\pi} }{{{\boldsymbol{\alpha }}^{T}}{{\mathbf{A}}_{b}}{{\boldsymbol{\alpha }}^{\text{*}}} \sin \tilde{\theta} d\tilde{\theta }}d{{\theta }_{p}}}}_{{\rho_{2}}} },
\end{align}
and
\begin{align}\label{MSE0_pULA}
{{\text{MSE}}_{0}}\left\{ {{{\tilde{f}}}_{d}} \right\} & \approx \frac{9} {\left( 2{\rm{\pi}}{\rho_{1}} \right)^2 }  \underbrace{\left[ \int_{0}^{ \rm{\pi} }\int_{0}^{ \rm{\pi} } \frac{10 - {{\left( { \rm{\pi} } {{f}_{d}}\tilde{x} \right)}^{2}}}{30} \sin \left( { \rm{\pi} }{{f}_{d}}\tilde{x} \right){{\boldsymbol{\alpha }}^{T}}{{\mathbf{A}}_{b}}{{\boldsymbol{\alpha }}^{\text{*}}}\sin 2\tilde{\theta }d\tilde{\theta }d{{\theta }_{p}} \right]^2}_{ \left( \varrho_{1} \right)^2}, \nonumber \\
{{\text{MSE}}_{0}}\left\{ {{{\tilde{\xi}}}} \right\} & \approx  \frac{9} { \left( {\rm{\pi}}{\rho_{2}} \right)^2 }  \underbrace{\left[ \int_{0}^{ \rm{\pi} }\int_{0}^{ \rm{\pi} } \frac{10 - {{\left( { \rm{\pi} } {{f}_{d}}\tilde{x} \right)}^{2}}}{30} \sin \left( { \rm{\pi} }{{f}_{d}}\tilde{x} \right){{\boldsymbol{\alpha }}^{T}}{{\mathbf{A}}_{b}}{{\boldsymbol{\alpha }}^{\text{*}}}\sin \tilde{\theta }d\tilde{\theta }d{{\theta }_{p}} \right]^2}_{ \left( \varrho_{2} \right)^2}.
\end{align}

In the case of fully calibrated ULA, ${\rho_{1}}$ could be simplified as
\begin{align}\label{rho1}
 {{\left. {{\rho }_{1}} \right|}_{\boldsymbol{\alpha }=\mathbf{1}}}= & \int_{0}^{ \rm{\pi} }{\int_{0}^{ \rm{\pi} } {{{\cos }^{2}}\tilde{\theta }{{\left. \left( {{\boldsymbol{\alpha }}^{T}}{{\mathbf{A}}_{b}}{{\boldsymbol{\alpha }}^{*}} \right) \right|}_{\boldsymbol{\alpha }=\mathbf{1}}}\sin \tilde{\theta }d\tilde{\theta }}d{{\theta }_{p}}} \nonumber \\
= & \int_{0}^{ \rm{\pi} }\int_{0}^{ \rm{\pi} } {{\mathbf{a}}^{T}}\left( {\tilde{\theta }} \right) {{\mathbf{a}}^{*}}\left( {{\theta }_{p}} \right)  \underbrace{{{\mathbf{a}}^{T}}\left( {{\theta }_{p}} \right) {{\mathbf{a}}^{*}}\left( {\tilde{\theta }} \right) \frac{{{\cos }^{2}}\tilde{\theta } }{\sin {{\theta }_{p}}}}_{f\left( {\tilde{\theta }} \right)}d\cos \tilde{\theta }d\cos {{\theta }_{p}} \nonumber \\
\approx & \int_{0}^{ \rm{\pi} }{\frac{1}{\tilde{d}} {{\mathbf{a}}^{T}}\left( {{\theta }_{p}} \right) {{\mathbf{a}}^{*}}\left( {{\theta }_{p}} \right) \frac{{{\cos }^{2}}{{\theta }_{p}}}{\sin {{\theta }_{p}}}\sin {{\theta }_{p}}d{{\theta }_{p}}} =\frac{{ \rm{\pi} }M}{2\tilde{d}},
\end{align}
and in the same way, we get ${\left. {{\rho }_{2}} \right|}_{\boldsymbol{\alpha } = \mathbf{1}} = \int_{0}^{ \rm{\pi} }{\int_{0}^{ \rm{\pi} } {{{\left. \left( {{\boldsymbol{\alpha }}^{T}}{{\mathbf{A}}_{b}}{{\boldsymbol{\alpha }}^{*}} \right) \right|}_{\boldsymbol{\alpha }=\mathbf{1}}}\sin \tilde{\theta }d\tilde{\theta }}d{{\theta }_{p}}} \approx \frac{{ \rm{\pi} }M}{\tilde{d}}$.
Substituting the simplified ${\rho_{1}}$ and ${\rho_{2}}$ into (\ref{MSEn_pULA}) leads to
${{\text{MSE}}_{\mathrm{n}}}\left\{ {{{\tilde{f}}}_{d}} \right\} \approx \frac{3\sigma _{\mathrm{n}}^{2}}{{{ \rm{\pi} }^{2}}MN}$ and ${{\text{MSE}}_{\mathrm{n}}}\left\{ {\tilde{\xi }} \right\} \approx \frac{3\sigma _{\mathrm{n}}^{2}}{\text{2}{{ \rm{\pi} }^{2}}MN}$.

Moreover, in the case of fully calibrated ULA, ${\varrho}_{2}$ could be simplified as
\begin{align}\label{varrho2}
{{\left. {{\varrho }_{2}} \right|}_{\boldsymbol{\alpha }=\mathbf{1}}} = & \int_{0}^{ \rm{\pi} }\int_{0}^{ \rm{\pi} } \frac{10 - {{\left( { \rm{\pi} } {{f}_{d}}\tilde{x} \right)}^{2}}}{30} \sin \left( { \rm{\pi} }{{f}_{d}}\tilde{x} \right) {{\left. \left( {{\boldsymbol{\alpha }}^{T}}{{\mathbf{A}}_{b}}{{\boldsymbol{\alpha }}^{*}} \right) \right|}_{\boldsymbol{\alpha }=\mathbf{1}}} \sin \tilde{\theta }d\tilde{\theta }d{{\theta }_{p}}, \nonumber \\
= & \int_{0}^{ \rm{\pi} }\int_{0}^{ \rm{\pi} } {{\mathbf{a}}^{T}}\left( {\tilde{\theta }} \right) {{\mathbf{a}}^{*}}\left( {{\theta }_{p}} \right)  \underbrace{{{\mathbf{a}}^{T}}\left( {{\theta }_{p}} \right) {{\mathbf{a}}^{*}}\left( {\tilde{\theta }} \right) \frac{10 - {{\left( { \rm{\pi} } {{f}_{d}}\tilde{x} \right)}^{2}}}{30} \frac{ \sin \left( { \rm{\pi} }{{f}_{d}}\tilde{x} \right) }{\sin {{\theta }_{p}}}}_{f\left( {\tilde{\theta }} \right)}d\cos \tilde{\theta }d\cos {{\theta }_{p}} \nonumber \\
\approx & \int_{0}^{ \rm{\pi} }{\frac{1}{\tilde{d}} {{\mathbf{a}}^{T}}\left( {{\theta }_{p}} \right) {{\mathbf{a}}^{*}}\left( {{\theta }_{p}} \right) \frac{10 - {{\left( { \rm{\pi} } {{f}_{d}} { {{\left. {\tilde{x}} \right|}_{\tilde{\theta} = \theta_{p} }} } \right)}^{2}}}{30} \frac{ \sin \left( { \rm{\pi} }{{f}_{d}} { {{\left. {\tilde{x}} \right|}_{\tilde{\theta} = \theta_{p} }} } \right) }{\sin {{\theta }_{p}}} \sin {{\theta }_{p}}d{{\theta }_{p}}} = 0.
\end{align}

Similar to (\ref{varrho2}), we have ${{\left. {{\varrho }_{1}} \right|}_{\boldsymbol{\alpha }=\mathbf{1}}} \approx 0$.
Hence, there hold ${{\text{MSE}}_{0}}\left\{ {{{\tilde{f}}_{d}}} \right\} \approx 0$ and ${{\text{MSE}}_{0}}\left\{ {{{\tilde{\xi}}}} \right\} \approx 0$.

\vspace{-0.6em}
\section{Expectation of ${\bf R}_{1,l,p}$ with respect to ${\theta }_{l,p}$}{\label{Expectation}}
For ease of representation, we simplify ${{\mathbf{R}}_{1,l,p}}$ as ${{\mathbf{R}}_{1}}=\mathbf{R}\left( \theta  \right)\otimes \left( \mathbf{e}\left( {{f}_{d}}\cos \theta  \right){{\mathbf{e}}^{H}}\left( {{f}_{d}}\cos \theta  \right) \right)$ and $\theta_{l,p}$ as $\theta$. Then the $(p,q)$th element of the $(m,n)$th submatrix of ${{\mathbf{R}}_{1}}$ is given by
\begin{align}
{{\mathbf{R}}_{1, mn-pq}}& ={{e}^{\text{j}2\chi\left( m-n \right)\cos \theta }}{{e}^{\text{j} \frac{2{\rm\pi}}{N}\left( p-q \right){{f}_{d}}\cos \theta }} ={{e}^{\text{j}x\cos \theta }}, x=2\chi\left( m-n \right)+2{\rm\pi} \frac{p-q}{N}{{f}_{d}}.
\end{align}

As $\theta \sim U\left(0,2\pi \right)$, we have
\begin{align}\label{R1Expectation}
  E\left\{ {{e}^{\text{j}x\cos \theta }} \right\} = \int_{0}^{2{\rm\pi}}{{\frac{{e}^{\text{j}x\cos \theta }}{2{\rm\pi}} }d\theta } =\int_{0}^{2{\rm\pi}}{\frac{\cos \left( x\cos \theta  \right)}{2{\rm\pi}} d\theta } = \frac{1}{2{\rm\pi}}\int_{-{\rm\pi}}^{{\rm\pi}}{\cos \left( x\sin \theta  \right)d\theta }={{J}_{0}}\left( x \right).
\end{align}

Define the operator $\oplus $ such that the $\left( m, n \right)$th submatrix of $A\oplus B$ is given by ${{a}_{mn}}+B$, where ${{a}_{mn}}$ is the $(m,n)$th element of $A$. Define ${{\mathbf{U}}_{M}}\in {{\mathbb{C}}^{M\times M}}$ whose $(m,n)$th element is $m-n$ and define ${{\mathbf{U}}_{N}}\in {{\mathbb{C}}^{N\times N}}$ whose $(p,q)$th element is $p-q$. Then with the results in (\ref{R1Expectation}), we readily arrive at (\ref{R1}), where $\mathbf{U}\left( {{f}_{d}} \right)$ is defined as $\mathbf{U}\left( {{f}_{d}} \right) = 2{\rm\pi}\frac{d}{\lambda }{{\mathbf{U}}_{M}} \oplus  2{\rm\pi} \frac{{{f}_{d}}}{N}{{\mathbf{U}}_{N}}$.

\vspace{-0.8em}
\section{Calculation of The First-Order Derivatives}{\label{Derivative}}
For zero-order Bessel function, there holds $\frac{\partial {{J}_{0}}\left( x \right)}{\partial x} =-\frac{1}{2}\left( {{J}_{1}}\left( x \right)-{{J}_{-1}}\left( x \right) \right)$.
Consequently, the first-order derivative (FOD) of ${{\mathbf{\tilde{R}}}_{1}}$ with respect to ${{f}_{d}}$ is given by
\begin{align}
 & \frac{\partial {{{\mathbf{\tilde{R}}}}_{1}}}{\partial {{f}_{d}}} =\frac{\partial {{J}_{0}}\left( \mathbf{U}\left( {{f}_{d}} \right) \right)}{\partial \mathbf{U}\left( {{f}_{d}} \right)}\odot \frac{\partial \mathbf{U}\left( {{f}_{d}} \right)}{\partial {{f}_{d}}} = -\frac{{{J}_{1}}\left( \mathbf{U}\left( {{f}_{d}} \right) \right)-{{J}_{-1}}\left( \mathbf{U}\left( {{f}_{d}} \right) \right)}{2} \odot \left( {{\mathbf{0}}_{M}}\oplus \frac{2{\rm\pi}}{N}{{\mathbf{U}}_{N}}  \right).
\end{align}

The FOD of ${{\mathbf{R}}_{2}}$ with respect to $\xi $ is given by
\begin{align}
 \frac{\partial {{\mathbf{R}}_{2}}}{\partial \xi } ={{\mathbf{1}}_{M}}\otimes \left( \mathbf{De}\left( \xi  \right){{\mathbf{e}}^{H}}\left( \xi  \right)+\mathbf{e}\left( \xi  \right){{\mathbf{e}}^{H}}\left( \xi  \right){{\mathbf{D}}^{H}} \right).
\end{align}

Define ${{\mathbf{i}}_{k}}\in {{\mathbb{C}}^{K\times 1}}$ as the $k$th column of the $K \times K$ identity matrix ${{\mathbf{I}}_{K}}$. Then the FOD of ${{\mathbf{R}}_{3}}$ with respect to the $k$th element of $\Re \left( \boldsymbol{\alpha } \right)$ and that of $\Im \left( \boldsymbol{\alpha } \right)$ could be respectively computed as
\begin{align}
\frac{\partial {{\mathbf{R}}_{3}}}{\partial \Re \left( {{\alpha }_{k}} \right)} =\left( \mathbf{U}\left( {{\mathbf{i}}_{k}}{{\boldsymbol{\alpha }}^{H}}+{\boldsymbol{\alpha}}\mathbf{ i}_{k}^{T} \right){{\mathbf{U}}^{T}} \right)\otimes {{\mathbf{1}}_{N}}, \ \frac{\partial {{\mathbf{R}}_{3}}}{\partial \Im \left( {{\alpha }_{k}} \right)} =\left( \mathbf{U}\left( \text{j}{{\mathbf{i}}_{k}}{{\boldsymbol{\alpha }}^{H}}-\text{j}{\boldsymbol{\alpha}}\mathbf{ i}_{k}^{T} \right){{\mathbf{U}}^{T}} \right)\otimes {{\mathbf{1}}_{N}}.
\end{align}

At last, $\frac{\partial \mathbb{R}}{\partial {{\eta }_{l}}}$ can be expressed as
\begin{align}
 & \frac{\partial \mathbb{R}}{\partial {{f}_{d}}}=\frac{\partial {{{\mathbf{\tilde{R}}}}_{1}}}{\partial {{f}_{d}}}\odot {{\mathbf{R}}_{2}}\odot {{\mathbf{R}}_{3}}\odot {{{\mathbf{\tilde{R}}}}_{4}},\
\frac{\partial \mathbb{R}}{\partial \xi } ={{{\mathbf{\tilde{R}}}}_{1}}\odot \frac{\partial {{\mathbf{R}}_{2}}}{\partial \xi }\odot {{\mathbf{R}}_{3}}\odot {{{\mathbf{\tilde{R}}}}_{4}},\ \frac{\partial \mathbb{R}}{\partial {{\sigma }_{\mathrm{n}}^{2}}} ={{\mathbf{I}}_{MN}}, \nonumber \\
 & \frac{\partial \mathbb{R}}{\partial \Re \left( {{\mathbf{\alpha }}_{k}} \right)} ={{{\mathbf{\tilde{R}}}}_{1}}\odot {{\mathbf{R}}_{2}}\odot \frac{\partial {{\mathbf{R}}_{3}}}{\partial \Re \left( {{\mathbf{\alpha }}_{k}} \right)}\odot {{{\mathbf{\tilde{R}}}}_{4}},\
\frac{\partial \mathbb{R}}{\partial \Im \left( {{\mathbf{\alpha }}_{k}} \right)} ={{{\mathbf{\tilde{R}}}}_{1}}\odot {{\mathbf{R}}_{2}}\odot \frac{\partial {{\mathbf{R}}_{3}}}{\partial \Im \left( {{\mathbf{\alpha }}_{k}} \right)}\odot {{{\mathbf{\tilde{R}}}}_{4}}.
\end{align}

\linespread{1.24}

\end{document}